\documentclass[10pt,conference]{IEEEtran}
\usepackage{cite}
\usepackage{amsmath,amssymb,amsfonts}
\usepackage{algorithmic}
\usepackage{graphicx}
\usepackage{textcomp}
\usepackage{xcolor}
\usepackage[hyphens]{url}
\usepackage{fancyhdr}
\usepackage{hyperref}

\usepackage{algorithm}
\usepackage{marginnote}
\usepackage{tcolorbox}
\usepackage{ulem}

\usepackage{microtype}
\usepackage{subfigure}
\usepackage{multicol}
\usepackage{multirow}
\usepackage{booktabs} % for professional tables
\usepackage{enumitem}

\definecolor{darkblue}{rgb}{0, 0, 0.5}
\definecolor{lightblue}{rgb}{0.68, 0.85, 0.9}

\newcommand{\diag}{\mathrm{diag}}

\newcommand{\vQ}{\mathbf{Q}}
\newcommand{\vK}{\mathbf{K}}
\newcommand{\vV}{\mathbf{V}}

\newcommand{\vS}{\mathbf{S}}

\newcommand{\vO}{\mathbf{O}}

\title{LeanAttention: Hardware-Aware Scalable Attention Mechanism for the Decode-Phase of Transformers}
\author{
    Rya Sanovar\\
    \and
    Srikant Bharadwaj\\
    \and
    Renee St. Amant\\
    \\Microsoft\\
    \and
    Victor Rühle\\
    \and
    Saravan Rajmohan\\
}
\begin{document}
\maketitle
\pagestyle{plain}

\begin{abstract}
Transformer-based models have emerged as one of the most widely used architectures for natural language processing, natural language generation, and image generation. The size of state-of-the-art models has increased steadily reaching billions of parameters. These huge models are memory hungry and incur significant inference latencies even on cutting edge AI-accelerators, such as GPUs. Specifically, the time and memory complexity of the attention operation is quadratic in terms of the total context length, i.e., prompt and output tokens. Thus, several optimizations such as key-value tensor caching and FlashAttention have been proposed to deliver the low latency demands of applications relying on such large models. However, these techniques do not cater to the computationally distinct nature of different phases during inference.

To that end, we propose LeanAttention, a scalable and generalized attention partitioning mechanism for the decode-phase of transformer-based models. LeanAttention enables scaling the attention mechanism for the challenging case of long context lengths by re-designing the attention execution flow for the decode-phase. We identify and prove the associative property of the softmax re-scaling operator, which allows it to function as a reduction operator. This property enables us to extend a "stream-K"-style reduction of tiled calculation to self-attention, which efficiently parallelizes attention computation over large context lengths and achieves near-perfect hardware occupancy irrespective of context size. As a result, we achieve an average of 1.73x speedup in attention execution compared to FlashDecoding, with up to 2.18x speedup for 256k context length.
\end{abstract}

\section{Introduction}
\label{sec:introduction}

Transformer-based~\cite{vaswani2023attention} language models~\cite{openai2024gpt4, touvron2023llama, zhang2022opt, li2023textbooks, chowdhery2023palm} have revolutionized the field of natural language processing (NLP) and found applications across diverse domains~\cite{microsoftcopilot, googlebard}. These powerful models, fueled by massive amounts of data and sophisticated architectures, have become indispensable tools for tasks such as machine translation~\cite{devlin2019bert}, question answering~\cite{chatGPT}, text generation~\cite{chatGPT}, and sentiment analysis. 

The core of the transformer architecture is the powerful component of self-attention. However, execution of the self-attention mechanism is slow and suffers from a large memory footprint, especially when dealing with longer contexts. A standard implementation of self-attention has quadratic time and memory complexity with respect to total sequence length, which leads to scalability challenges as model sizes~\cite{brown2020language} and supported context lengths increase~\cite{claude3,zhang2024soaring,liu2023ring}. 
Despite these scalability challenges, we see a trend of state-of-the-art models supporting greater and greater context lengths, with some production models supporting contexts hundreds of thousands of tokens long. Support for long context lengths can improve a model's utility by allowing for an increasingly rich context, which is particularly beneficial in a range of applications (e.g. RAG involving numerous or long documents) allowing improved relevance, coherence, and user experience.

To mitigate LLM scalability challenges, mechanisms like FlashAttention~\cite{dao2022flashattention} and FlashAttention-2/3~\cite{dao2023flashattention2, flashattention3} have been developed.
% FlashAttention employs an IO-aware exact attention algorithm that uses incremental softmax computation to minimize memory reads and writes to and from GPU high bandwidth memory.
FlashAttention brings IO-awareness to attention computation by reducing slow reads and writes to and from the GPU's global memory~\cite{ivanov2021data}. Instead, it computes attention in the faster shared memory using a tiling strategy. It allows for parallelization over batch size and number of heads. FlashAttention-2 builds on FlashAttention to further optimize attention computation by enabling parallelization over input sequence length (or query length). 
% Furthermore, its latest variant is FlashAttention-3~\cite{flashattention3}, which adopts an asynchronous low-precision execution of FlashAttention-2 designed to exploit the unique hardware capabilities (i.e. the TMA unit - Tensor Memory Accelerator) specific to the Hopper GPU~\cite{luo2024benchmarkingdissectingnvidiahopper}.
% additionally reducing the number of non-matrix multiply operations to maximize GPU throughput, and it additionally enables parallelization across input sequence length (query length) as well.

While these optimizations provide significant improvements,
% (e.g. FlashAttention-3 being 1.5-2x faster than FlashAttention-2 which itself realizes 2x speedup over FlashAttention), 
these mechanisms only provide performance benefits for a subset of problem sizes (i.e. query length, context length, batch size, and number of heads). Their utilization of underlying hardware resources is mostly optimized for problem sizes encountered in the prefill-phase of transformer-based models, and often results in critically low hardware utilization for problem sizes found in the decode-phase. 
By overlooking the distinct behavior of attention during the decode phase versus the prefill phase, these mechanisms miss out on potential performance gains that could be achieved by efficiently exploiting the parallelization capabilities of the underlying hardware.

% As context lengths grow, its evident that parallelization of attention computation over the context length dimension is severely necessary. Although mechanisms like FlashDecoding~\cite{flashdecoding} and FlashInfer\cite{flashinfer} enable this parallelization through the fixed-split partitioning strategy, they still suffer from hardware resource underutilization and unnecessary overheads, where both are highly contingent on problem size.

In decoder-only transformer models, the inference process for a single request involves multiple forward passes through the model where output tokens are generated sequentially~\cite{fullstackoptimization}. This inference procedure inherently comprises of two distinct computational phases due to the practice of reusing (caching) the key-value tensors of the previously computed tokens~\cite{pope2023efficiently}.
The first phase is the \textit{prefill-phase} (sometimes known as \textit{prompt-computation phase}) where attention is computed of the entire input prompt against itself to generate the first output token. 
This phase is computationally intensive and demands high FLOPS/s (floating point operations per second)~\cite{fullstackoptimization}. 
Following the prefill-phase, the \textit{decode-phase} (sometimes known as \textit{token-generation phase}) begins in an auto-regressive manner~\cite{vaswani2023attention}. Each subsequent token is produced based on the attention computed of the preceding token against itself and the entire cached context (\textit{kv-cache}) of previous tokens in the sequence. 
% -- commented line out for restructuring. move to background section --
%\RSTA{Maybe move down: This cached context ($N_k$) is usually very long, exceeding more than a million tokens in length, whereas the query length ($N_q$) is 1 for every autoregressive step.} 
With the push towards longer contexts, this cached context length can get extremely long, exceeding more than hundreds of thousands of tokens in length~\cite{li2024longcontext,zhang2024soaring,fu2024data,liu2023ring}. Despite state-of-the-art batching techniques~\cite{yu2022orca} and attention partitioning mechanisms~\cite{dao2022flashattention, dao2023flashattention2, flashdecoding, flashinfer}, the lack of a smart parallelized execution of attention along this long context length makes the decode-phase slow, bound by memory bandwidth ~\cite{williams2009roofline} and capacity ~\cite{fullstackoptimization}. Importantly, as we discuss in \autoref{sec:challenges}, even when the query length is significantly longer than the total number of output tokens produced during inference, the majority of the overall processing time is consumed by the decode-phase.

% Use figure* and \textwidth instead of \columnwidth to set the size of the
% figure to the whole width of the text box (two columns).
\begin{figure}[t]
\vspace{-0.1in}
\begin{center}
\centerline{\includegraphics[width=0.7\columnwidth]{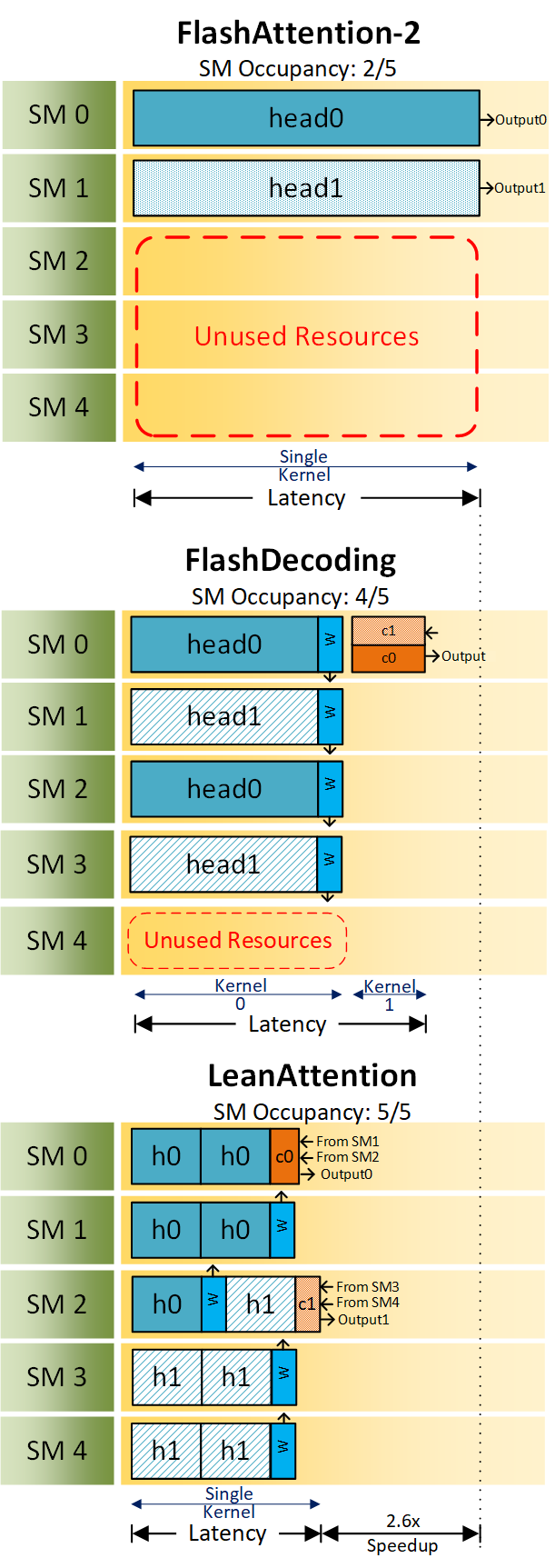}}
\vspace{-0.1in}
\caption{Execution schedule of FlashAttention-2\cite{dao2023flashattention2}, FlashDecoding\cite{flashdecoding} and FlashInfer\cite{flashinfer} (fixed-split), and LeanAttention across a hypothetical five SM GPU executing attention of 2 heads. LeanAttention splits the context into optimal LeanTiles (shown here with 5 tiles per head).}
\label{fig:mainfig}
\end{center}
\vskip -0.2in

\end{figure}

It's evident now that efficient parallelization of attention computation over the context length dimension is highly necessary. Although mechanisms like FlashDecoding~\cite{flashdecoding} and FlashInfer\cite{flashinfer} enable this parallelization through the fixed-split partitioning strategy, they provide the hardware with imbalanced loads and as a consequence still suffer from hardware resource underutilization and unnecessary reduction overheads -  both being highly contingent on problem size. Further, attention optimizations~\cite{sarathi} are increasingly relying on batching requests of unequal context lengths to improve overall throughput but suffer from hardware under-utilization because of the partitioning strategies adopted by FlashAttention and FlashDecoding/FlashInfer.

In this work, we aim to address the limitations of previous work as it relates to the decode-phase of inference, which we find exhibits unique computational characteristics in comparison to the prefill-phase. 

% In summary, for the decode phase of LLM inference, FlashAttention-2 parallelizes computation only along the number of heads in the model and batch size. Although FlashDecoding~\cite{flashdecoding} and FlashInfer\cite{flashinfer} enable parallelization along context length through the method of fixed-split partitioning, they still suffer from inefficient load balancing, hardware resource underutilization, and unnecessary overheads that scale with problem size. 

We introduce \emph{LeanAttention}, a generalized \emph{exact-attention} mechanism which enables parallelization across all problem size dimensions, ensures perfect quantization efficiency, i.e. 100\% GPU occupancy, for all problem sizes with constant reduction overheads, delivers a runtime speedup in attention computation for long context lengths, and is scalable to multi-GPU scenarios with tensor parallelism. 

% As shown in \autoref{fig:mainfig}, we aim to execute attention in a single fused kernel with a greater degree of parallelization, i.e. across the context length as well \textit{while} ensuring peak processor occupancy and minimal reduction overheads that are independent of problem size (in both prefill and decode phase). 

Overall, our contributions are as follows:
\begin{itemize}[noitemsep]

    \item Identify the limitations of state-of-the-art attention execution optimizations on GPUs during the decode-phase of transformer-based models. (\autoref{subsec:limitations})
%%%%%%%%%%%%%% REMOVED %%%%%%%%%%%%%%
    % \item Approach the softmax operation in attention as a special reduction operation and extract it out of the inner loop of the attention algorithm, thus treating softmax re-scaling of partial attention output tensors as an \textit{associative reduction operation}. (\autoref{subsec:softmaxreduction})

%%%%%%%%% NEW %%%%%%%%%%%%%%
    \item Identify the \textit{associative nature} of the softmax re-scaling operator that enables it to function as a reductive operator, and leverage this crucial property in LeanAttention to split an independent attention workload into unequal sizes along the context length dimension when needed. (\autoref{subsec:softmaxreduction})
    \item Leverage a stream-K style~\cite{streamk} partitioning in LeanAttention that \textit{always} provides \textit{equalized} compute loads to every compute unit in the hardware system (as shown in \autoref{fig:mainfig}), thus giving near 100\% hardware occupancy and delivering speedup irrespective of problem size and hardware architecture. (\autoref{subsec:streamkmapping})
    % \item Define LeanAttention as a generalized attention partitioning mechanism that closely mirrors attention computation to the compute and memory hierarchy of modern hardware systems and that is optimized for decode-phase problem sizes in addition to prefill-phase. (\autoref{sec:Method})
    \item Expatiate LeanAttention's versatility and generalizability, where FlashAttention-2 and FlashDecoding (as well as FlashInfer) can be recovered as special cases of it. (\autoref{subsec:streamkmapping})

\end{itemize}
As detailed in \autoref{sec:evalresults}, LeanAttention results in an average of 1.73x latency speedup over FlashDecoding for the decode phase of transformer-based models and up to 2.18x speedup over FlashDecoding for 256k context length, while maintaining a near 100\% GPU occupancy irrespective of problem size. 

% \RS{Don't need:
% As such, LeanAttention not only enables speedups with its efficient load balancing strategy which proves to be very useful for decoding long context lengths, it changes the architectural design tradeoffs and opens the door to a new line of microarchitecture evolution that leverages this newly enabled parallelism and compute utilization.
% }

% \begin{figure}[ht]
% \begin{center}
% \centerline{\includegraphics[width=\columnwidth]{Figures/flashattnparta.png}}
% \vspace{-0.3in}
% \caption{Iterative update of output in FlashAttention-2 with $C_n = 3$ iterations. }
% \label{fig:flashattnparta}
% \end{center}
% \vskip -0.2in

% \end{figure}
\section{Background}
\label{sec:background}

In this section, we provide required background on Standard Attention \cite{vaswani2023attention} and FlashAttention-2 \cite{dao2023flashattention2}. 

\subsection{Standard Attention}
% We will first explain some elementary details of the standard attention implementation.
For a given input tensor with dimensions of batch size $B$, query length $N_q$, key/value sequence length (also known as context length) $N_k$, and hidden dimension $D$, multi-head attention typically splits attention computation into $h$ number of heads along the hidden dimension, with each head responsible for computing attention independently for a head dimension of size $d = D/h$. 

Unlike standard transformer execution, the query length and context lengths may not always be equal, where key-value tensors are cached~\cite{pope2023efficiently}. For instance, the prefill-phase of generative decoder-only transformers such as GPT-4~\cite{openai2024gpt4} or Phi-2~\cite{li2023textbooks} has sequence lengths $N_q = N_k = N$, but in their decode-phase the context length increments by 1 after every autoregressive step of decode generation, while the query length, for a given batch instance and head, is the singular token that was generated in the previous $n$-th time step, i.e., $N_q = 1$ and $N_k = N + n$.

The query matrix $Q \in R^{N_q \times d}$ and key $K$ and value $V$ matrices $\in R^{N_k \times d}$ are inputs to the following equation which is computed independently by the different batch instances and heads. The output matrix $O \in R^{N_q \times d}$ is obtained in essentially three steps as shown in Equation \ref{eqn:stdattn}. Table~\ref{tab:dimtable} summarises the three operations involved in self-attention along with their corresponding dimensions in both prefill and decode-phase at time step $n=0$.

\begin{equation}
    S = QK^T,\text{ } P = softmax(\frac{S}{\sqrt{d}}),\text{ } O = PV
    \label{eqn:stdattn}
\end{equation}

% Standard attention implementation involves computing the large intermediate matrices, namely the attention score matrix $S \in R^{N_q \times N_k}$ and the softmax matrix $P\in R^{N_q \times N_k}$ and storing them in global memory. It requires \textit{a priori} knowledge of all tokens in a row of the attention score matrix for computing the softmax function. Specifically, the row-wise maximum and exponential sum must be computed beforehand, which requires examining all tokens in a given row of the attention score matrix and thus necessitates storing the intermediate $S$ and $P$ matrices.

Standard attention implementation involves computing the large intermediate matrices, namely the attention score matrix $S \in R^{N_q \times N_k}$ and the softmax matrix $P\in R^{N_q \times N_k}$ and storing them in global memory. These intermediate matrices need to be stored in the global memory because the computation of the softmax matrix $P$ requires \textit{a priori} knowledge of all tokens in a given row, specifically the row-wise maximum and exponential sum of tokens in a row need to be computed beforehand to calculate the softmax-ed value of each element in the row.

The computational complexity of standard attention is on the order of $O(N_q N_k d)$, with the two matrix multiplications (MatMul's) contributing to the majority of it. Due to slow global memory access speeds, storing and retrieving these intermediate matrices ~\cite{ivanov2021data} is costly in terms of latency and incurs a large memory footprint, both in the order of $O(N_q N_k)$. 

\begin{table}[htb]
    \centering
    \resizebox{\columnwidth}{!}{
    \begin{tabular}{|c|c|c|c|}
\hline \multirow{2}{*}{ Operation } & \multirow{2}{*}{ Type } & \multicolumn{2}{|c|}{ Operation Dimension } \\
\cline { 3 - 4 } & & Prefill & Decode \\
\hline $query \times key$ & MatMul & $N \times d \times N$ & $1 \times d \times N$ \\
\hline $softmax$ & EleWise & $N \times N$ & $1 \times N$ \\
\hline $attn\_score \times value$ & MatMul & $N \times N \times d$ & $1 \times N \times d$ \\
\hline
\end{tabular}
    }
    \caption{Operations in Self-Attention. Matrix multiplications are described in the $M \times N \times K$ format.}
    \label{tab:dimtable}
\end{table}

\subsection{Flash Attention-2}
\label{subsec:FA2background}
To mitigate the memory footprint and access overhead ~\cite{ivanov2021data} associated with storing the $S$ and $P$ matrices, FlashAttention employs kernel fusion of the three operations as shown in \autoref{eqn:stdattn}: $query \times key$ MatMul, softmax and $attn\_score \times value$ MatMul, requiring no intermediate global memory reads and writes. To this end, it employs the tiling strategy.

By utilizing the online softmax algorithm~\cite{milakov2018online}, FlashAttention only requires a single pass over an entire row of tokens to compute their softmax, bypassing the issue of \textit{a priori} knowledge in standard attention. This helps leverage the tiling strategy which partitions the attention output matrix $O$ into independent output tiles (attention computation an output tile is independent of the computation of other output tiles). A grid of \textit{cooperative thread arrays} (CTAs)\footnote{Blocks of GPU threads are coscheduled in CTAs, which virtualize the hardware's streaming multiprocessor cores (SMs)} is launched, each computing a given output tile of the output matrix $O$. The input matrices $Q$, $K$ and $V$ are partitioned into smaller tiles too. While computing the output tile corresponding to a given query tile, the key/value tiles are brought into shared memory in a \textit{sequential} manner and the attention output tile is continuously updated and corrected by the right scaling factors. This on-chip updation avoids the need of storing the intermediate $S$ and $P$ matrices in global memory. In addition to parallelizing computation over batches and heads like FlashAttention, FlashAttention-2 further parallelizes over the query length dimension, as the attention computation of output tiles along this length is independent. This results in a 2x speedup over FlashAttention.

Thus, the tiling strategy ensures that the extra global memory space required by FlashAttention-2 is $O(N_q)$ (needed to store the logexpsum $L$ for backward pass), an impressive improvement in memory footprint over the $O(N_q \times N_k)$ in traditional attention. The additional parallelism over query length helps it reach 50-70\% of peak theoretical FLOPS/s and increases hardware occupancy in the prefill phase. FlashAttention-2 was augmented to FlashAttention-3~\cite{flashattention3}, specifically fine-tuned for execution on Hopper H100 GPU's~\cite{luo2024benchmarkingdissectingnvidiahopper} to exploit its low-precision and asynchronous hardware capabilities. 

Like FlashAttention-2, other related techniques such as Ring Attention\cite{liu2023ring} and Striped Attention\cite{brandon2023striped}, are optimized for prefill-phase problem sizes and thus suffer from longer latencies during the decode phase. 

\section{Challenges In The Decode Phase}
\label{sec:challenges}
Prior to outlining our methodology for LeanAttention, to set the stage for our approach, we delve into some of the challenges encountered in the decode phase of LLM inference, as well as the limitations of FlashAttention-2 optimizations in the decode phase.

\subsection{Time Spent in Decode Phase}
\label{subsec:decodetimespent}

As we've discussed, modern generative LLM inference comprises of two computationally distinct phases: the prefill phase followed by the decode phase. In the prefill phase, self-attention is computed for the entire input prompt. The query length, $N_q$, in this phase is the same as the context length, $N_k$, i.e., ($N_q = N_k = N$). Whereas, the decode phase begins generating each subsequent output token in autoregressive iterations. For each iteration of the decode phase, its query length is a single token, $N_q=1$, and its context length, $N_k$, could be very long, in the order of more than thousands of tokens depending on the auto-regressive step and input query length.

% Generative LLM inference comprises two distinct computational phases: the prompt computation phase (sometimes called the prefill phase) and the decode phase (sometimes called the token generation phase). In the prefill phase, all tokens in the input prompt (aka query) undergo parallel forward passes through the model to generate the first output token. 
% The query length, $N_q$, in this phase is the same as the context length, $N_k$, resulting in an $N \times N$ attention matrix ($N_q = N_k = N$). This phase is computationally intensive and demands high FLOPS/s. 

% Following the prefill phase, the decode phase begins generating each subsequent output token. The nature of this phase is auto-regressive, where the next output token is produced based on the forward pass of the preceding token and the cached context (KV cache) from previous tokens in the sequence. For each iteration of the decode phase, its query length is a single token, $N_q=1$, and its context length, $N_k$, could be in the order of more than thousands of tokens depending on the auto-regressive step and input query length. 

\begin{figure}[!t]
\begin{center}
\centerline{\includegraphics[width=\columnwidth]{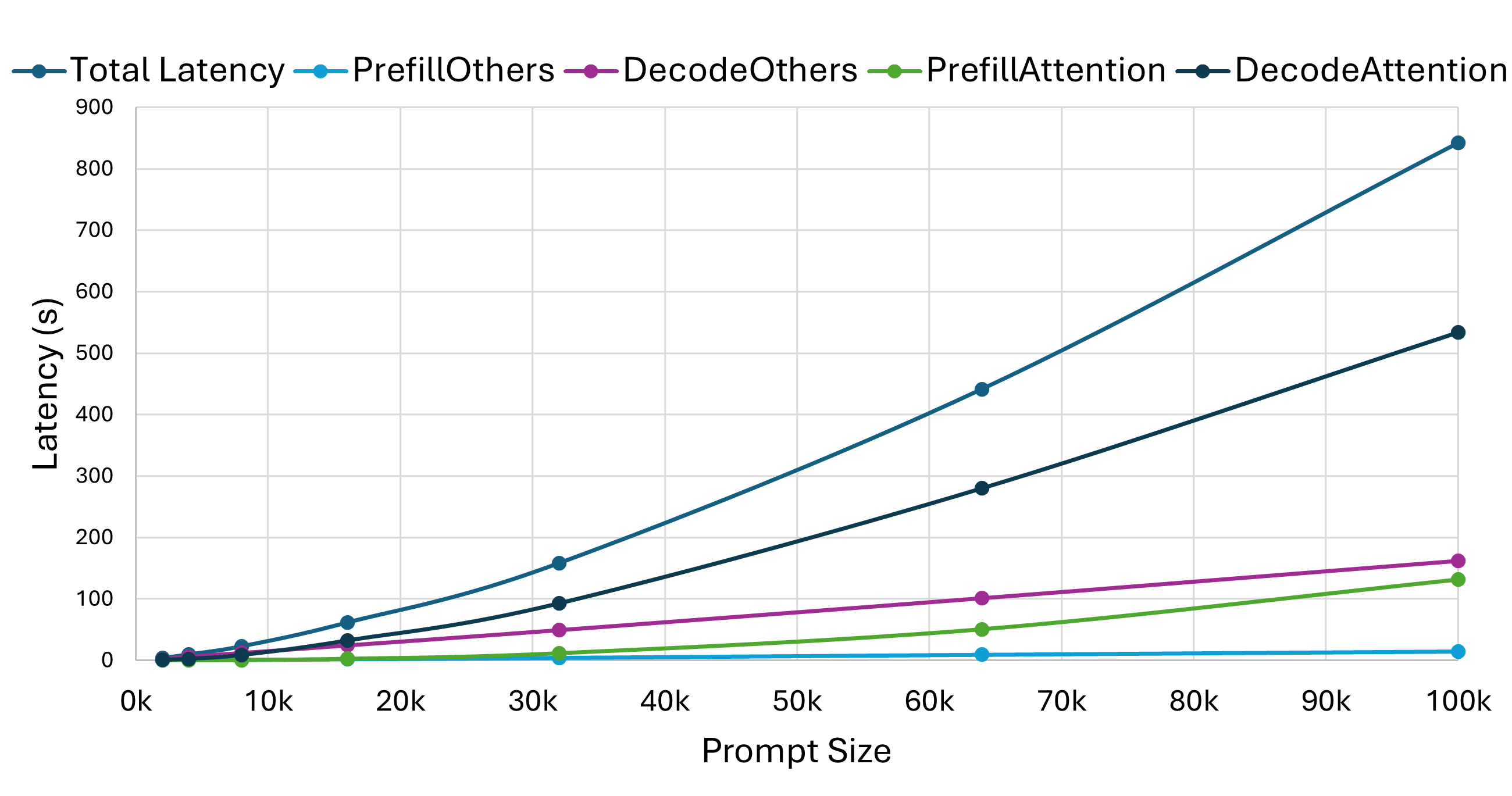}}
\vspace{-0.1in}
\caption{Timeshare of decode attention compared to other stages for different prompt sizes with 8:1 token ratio for Phi-3 Medium model with single batch size.}
\label{fig:token_timeshare}
\end{center}
\vspace{-0.2in}

\end{figure}

\autoref{fig:token_timeshare} depicts the processing time breakdown of the prefill and decode phase, with the decode phase's further breakdown into time spent in the $Q/K/V$ activation layer, the decode attention layer and the feed-forward linear layers.  

While the large matrix multiplications found in the linear layers of the prefill phase of inference are heavily optimized (all the model layers during prefill phase only taking up 10\% of the timeshare even for a high prompt:output ratio), the decode phase presents a different challenge. 
During the decode phase, where the query length is only 1 token long, linear layers perform matrix multiplications on very narrow matrices which do not provide enough work to occupy the GPU. MatMul partitioning strategies like Stream-K~\cite{streamk} can be leveraged to efficiently partition these narrow matrices and accelerate their computation, preventing the linear layers from becoming a bottleneck during decode phase. However, the attention layer, with the existing attention partitioning techniques~\cite{dao2022flashattention, dao2023flashattention2, flashdecoding, flashinfer} experiences longer latencies along with significant underutilization of hardware resources during this phase. 
This makes leveraging efficient parallelism along context length ($N_k$) during attention a crucial aspect in increasing SM occupancy and reducing decode phase processing time.

As the number of output tokens generated rises, the context length becomes longer and thus the proportion of time spent in the decode phase relative to the prefill phase becomes larger. \autoref{fig:token_timeshare} depicts this imbalance in processing time spent in the prefill phase and the decode phase attention. Even with a prompt input to output token ratio of 8:1, more than 50\% of the processing time is consumed by the decode phase, taking up to nearly 80\% of the timeshare for longer prompt sizes. Additionally, other layers of the decode phase such as QKV and MLP (FFN layers) can be optimized using state-of-the-art MatMul partitioning techniques such as Stream-k. These operations are typically quantized to lower data formats such as INT8 to further enhance their efficiency. As a result, the attention operation can constitute up to 40-50\% of the total duration of decode phase inference as shown.

\subsection{Limitations of FlashAttention-2 for Decode}
\label{subsec:limitations}

\begin{figure}[t]
\begin{center}
\centerline{\includegraphics[width=\columnwidth]{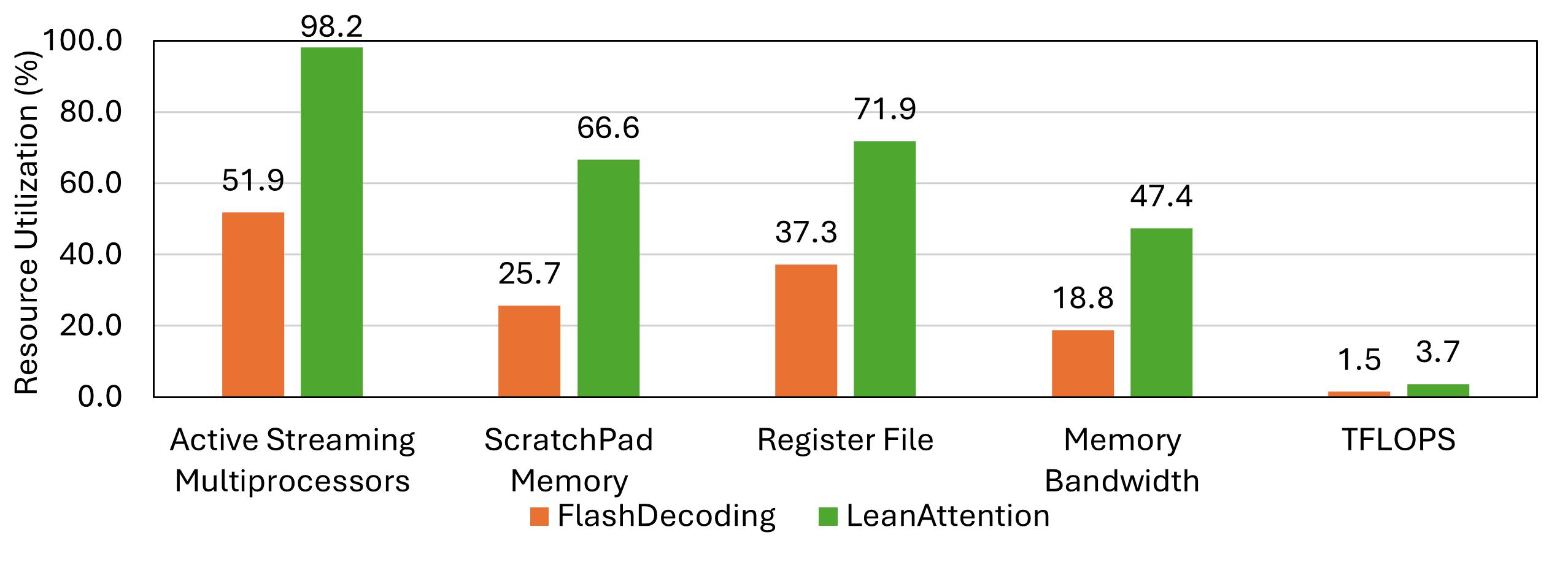}}
\vspace{-0.1in}
\caption{Utilization of various resources on a single Nvidia-A100-80GB GPU in LeanAttention compared to FlashDecoding kernel at Heads=56 and BS=1 measured using Nsight Compute. FlashDecoding has a quantization efficiency issue with the 108 SMs on the GPU. LeanAttention occupies all SMs available in the system.}
\label{fig:sm_occupancy}
\end{center}
\vspace{-0.2in}

\end{figure}

In both the prefill and decode phase, FlashAttention-2 traverses the context length dimension ($N_k$) sequentially, i.e., it updates the attention output for a given query tile by bringing in the key/value tiles into shared memory in a \emph{sequential} manner. While FlashAttention-2 does parallelize over query length ($N_q$) to increase SM occupancy, this additional mode of parallelism has limited parallelization capacity in the decode phase where the query is a single token ($N_q = 1$). Not parallelizing attention computation along context length makes FlashAttention-2~\cite{dao2023flashattention2} suffer from extremely low SM occupancy during decode as depicted in \autoref{fig:sm_occupancy}. This means that at any given point in time, the number of CTAs in flight on the GPU is directly proportional to the number of query tiles, and, therefore, to the query length - regardless of the context length.

More explicitly, for a single batch instance, the maximum number of heads for state of the art LLMs barely occupy the compute resources of modern hardware architecture systems during the decode phase where query length $N_q = 1$. For example, for a model with 128 heads, its decode phase would suffer from severe under-utilization of an 8 GPU A100 system that has 864 compute cores at its disposal. Unlike the prefill phase, decode phase can offer parallelization only across batch size and number of heads for FlashAttention-2.

Processor occupancy in FlashAttention-2 could be improved by increasing the batch size or number of heads, the other two modes of parallelization it addresses. Intuitively, having larger batch sizes in the decode phase could provide enough work to every compute resource to fully occupy the GPU, but this introduces other challenges and limitations. Due to increasingly large model sizes, the need to independently cache KV context for every batch instance would likely exceed the memory capacity of the hardware system. Moreover, scheduling overheads~\cite{orca22} for efficiently batching queries along with the challenges of batching low SLA queries would increase inference latency and challenge utilization. 

Without having to resort to larger batch sizes as the sole solution to resolving the GPU occupancy issue (which is limited by available memory capacity), the large context length in the decode phase would benefit from partitioning its workload across different SMs efficiently. This motivates the need for smarter attention decomposition techniques which can efficiently distribute the workload across the cores without resorting to larger batch sizes. 

\subsection{Limitations of Related Work}
\label{subsec:fixedsplit}

FlashDecoding, which is FlashAttention-2 with \emph{fixed-split} partitioning, has recently been proposed \cite{flashdecoding,hong2024flashdecoding,githubGitHubDaoAILabflashattention}, where attention computation is also partitioned along context length $N_k$. FlashInfer\cite{flashinfer} implements an identical fixed-split partitioning of attention for single-request's in decode phase in pure CUDA. For the case of batched-requests in decode, FlashInfer implements an optimized version of PagedAttention for efficient KV cache storing and fetching.

\emph{Fixed-split} is a general matrix multiplication decomposition scheme that we briefly describe here. 
% \RSTA{TODO DECIDE IF WE NEED MORE DETAIL IN AN Appendix}. 
Given a MatMul computation problem with matrices A $(M \times K)$ and B $(N \times K)$ to obtain a matrix C $(M \times N)$ where $C = AB^T$, to optimize concurrent computation, the fixed-split mechanism~\cite{cutlassgithub} partitions the K-mode of the A and B matrices into $s$ batches based on a fixed splitting factor $s$ provided dynamically at run time. 
This launches $s$ times the CTAs (Cooperative Thread Arrays, equivalent to a threadblock) as launched without fixed-split, which are computing partial products of the output tiles of the C matrix concurrently. Fixed-split utilizes the associativity of addition in the inner product of a MatMul to later reduce or ``fix-up" the partially computed C matrices to produce the final C matrix. The concurrency from fixed-split reduces latency and simultaneously increases hardware occupancy at the cost of an additional reduction at the end. 

%%%%%%%%%%%% REMOVED %%%%%%%%%%%%%%%
% It eliminates synchronization costs by approximating a global max value in softmax to avoid computing local softmax and final re-scaling. A global softmax is computed at the end once all the partial exponential sums have been determined. Additionally, double buffering is used to hide memory access latencies. 
FlashDecoding++~\cite{hong2024flashdecoding}  achieves speedup over FlashDecoding by approximating the softmax operation to remove the sequential dependencies it creates in attention. 
Notably, this approach compromises on accuracy and its implementation is limited to certain model architectures. In contrast, LeanAttention computes exact attention with no loss in accuracy and can be used in any transformer-based model. 
FlashDecoding++, as well as other works that focus on softmax approximations to achieve speedup (like ConSmax~\cite{53124} and Softermax~\cite{stevens2021softermax}), can be seamlessly integrated into LeanAttention.

Despite these improvements, fixed-split used in these mechanisms~\cite{flashdecoding, flashinfer, hong2024flashdecoding} is a non-optimal load balancing strategy. While this method of partitioning would provide speedup and occupy the GPU well for some attention workloads, it is an inefficient load balancing strategy for the entire problem space and often results in partially full waves of attention computation that suffer from quantization inefficiencies, i.e. low GPU occupancy due to imbalanced loads, and loses out on performance gains it could get from the idle resources otherwise (depicted in \autoref{fig:mainfig}). While increasing the number of splits could help occupy the GPU better, it would result in reduction overheads that scale with the split factor and would allocate minimal work to each SM making it an inefficient use of register space. 

This fixed-split partitioning along the context length does occupy a larger number of compute resources on the GPU compared to vanilla FlashAttention-2, but it often provides imbalanced loads to the compute units, and GPU occupancy then varies greatly with problem size, split factor and number of compute units in the hardware system as shown in \autoref{fig:sm_occupancy}, making it unlikely for FlashDecoding and its variants to reach perfect quantization efficiency for all problem sizes and hardware systems. Moreover, the problem of quantization inefficiencies with these mechanisms would be particularly exacerbated in the common cases of processing a batch of requests of heterogenous context lengths~\cite{sarathi}. In contrast, LeanAttention, with its stream-K-style decomposition discussed in \autoref{sec:Method}, will \textit{always} provide well-balanced loads to each compute unit in the hardware system and reach near 100\% GPU occupancy for all problem sizes and hardware architectures, making it perfectly adept to handle batched requests of unequally sized contexts. 

\subsection{Multi-GPU Execution with Tensor Parallelism}
FlashAttention-2 not only severely underutilizes GPU cores in the decode phase, but is also not adaptable to multi-GPU scenarios due to its lack of support for tensor parallelism. This makes FlashAttention-2 less scalable to multi-GPU systems which has become an imperative due to capacity-boundedness of contemporary large language models~\cite{brown2020language} and the support they require for increasingly long context lengths~\cite{claude3}. This asserts the need for an attention mechanism that also scales well to multi-GPU scenarios.

These challenges motivate the need for a generalized attention mechanism that works for a vast set of problem sizes (in both prefill and decode phase) and is closely aligned with the memory and compute hierarchies of modern hardware systems. We formulate this scalable and generalized exact attention mechanism as \textbf{LeanAttention}, which computes exact attention faster in a single fused kernel launch, has optimal quantization efficiency for all kinds of problem sizes, whilst also being scalable to multi-GPU scenarios through its support for tensor parallelism.
\section{LeanAttention}
\label{sec:Method}

\begin{figure}[t]
\begin{center}
\centerline{\includegraphics[width=\columnwidth]{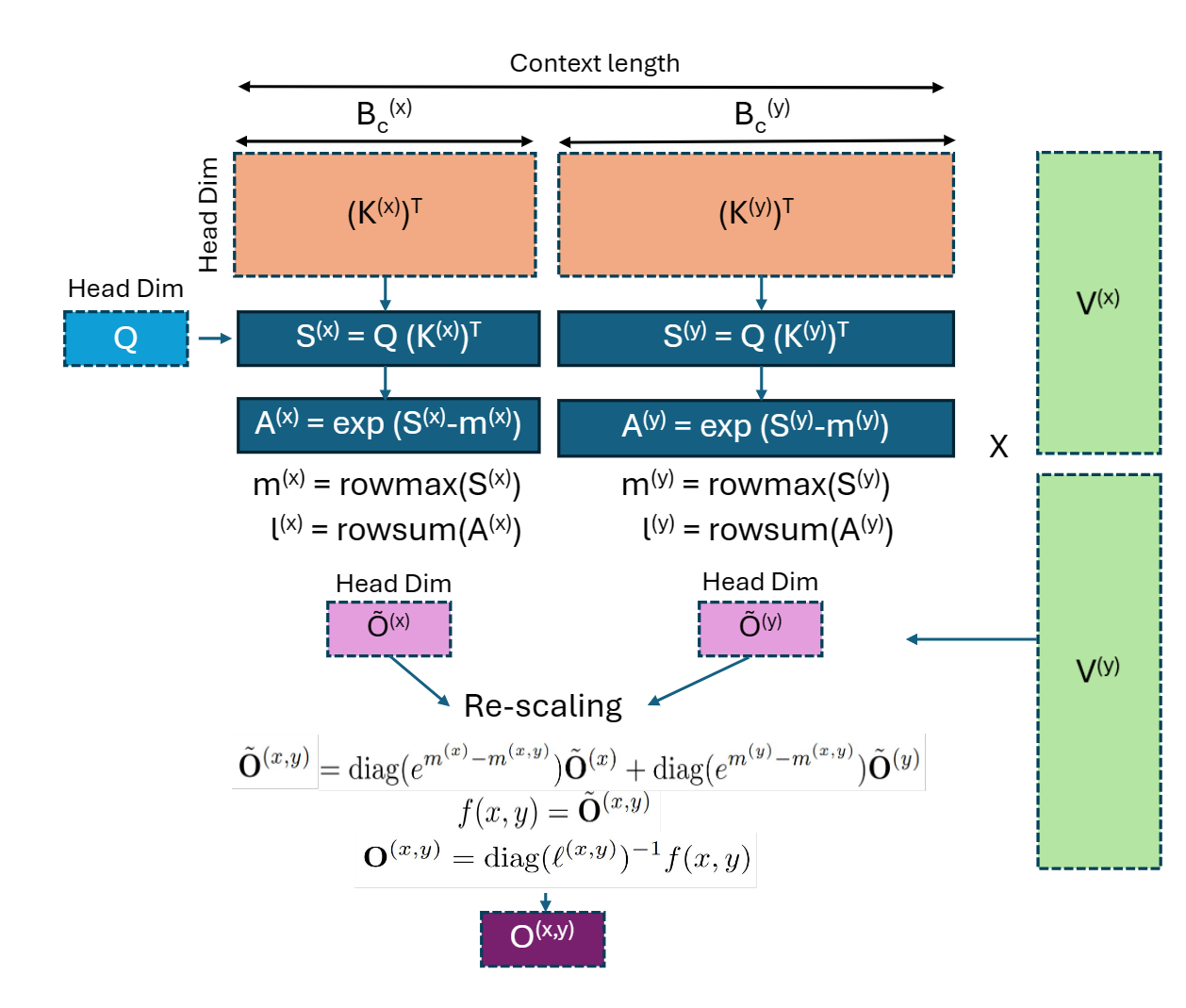}}
\vspace{-0.2in}
\caption{Illustrative diagram showing LeanAttention's partitioning strategy with two differently sized work volumes of a head assigned to different CTAs. The un-scaled outputs are independently computed and re-scaled later in a reduction operation. Note that this can be generalized to any arbitrary-sized work volume split.}
\label{fig:leanmath}
\end{center}
\vspace{-0.3in}
\end{figure}

% \begin{figure*}
% \includegraphics[width=0.9\textwidth]{Figures/LeanAttnFigFinal.png}
% \vspace{-0.2in}
% \caption{Algorithmic flow of LeanAttention for 2 heads in decode phase. Each head's output tile is decomposed into 5 LeanTiles.}
% \vspace{-0.2in}
% \label{fig:LAwithHeads}
% \end{figure*}

LeanAttention is an optimized scalable attention execution mechanism. It provides extensive parallelism across all modes of the attention tensor, with well-balanced computation workloads to each CTA ensuring close to 100\% SM occupancy while delivering a runtime speedup in attention execution.

% \RS{don't need:
% To exploit regular matrix multiplication optimizations on the GPU, such as the state-of-the-art ``stream-K" ~\cite{streamk} for the case of attention, we restructure attention to closely mirror the execution paradigms of a MatMul computation.}

% \textcolor{magenta}{
% The attention operation consists of two MatMuls joined by a softmax operation. For a stream-K style partitioning of these attention MatMuls on the GPU, we define the softmax re-scaling operation in attention as a final reduction operation to replace the addition operation that behaves as reduction in a basic MatMul.
% }

First, we identify the smallest optimal granularity of decomposition in attention computation, termed as \textit{LeanTile} (\autoref{subsec:leantile}), which can be linearly mapped on the hardware resources in a flexible style akin to stream-k decomposition of matrix multiplications \autoref{subsec:streamkmapping}). Multiple such LeanTiles belonging to either single or multiple attention outputs will constitute a workload assigned to a CTA. By the nature of Stream-K's equalized load balancing strategy, each CTA will compute equal number of LeanTiles, ensuring no idle SMs during the entire duration of attention computation.

Second, we identify that the \emph{associative property of softmax re-scaling} enables us to treat it as a reduction operation along the context length dimension and allows us to split the workload (i.e. KV tensors) of a single head into unequally sized blocks (described in \autoref{subsec:softmaxreduction}). In the most common case where processor width (number of SMs) is not a multiple of the total number of heads, we must split the workload of each head into unequal sizes on the SMs \textit{in order} for the \textit{total} workload per SM to be equal (unlike FlashDecoding, FlashInfer and variants), and by virtue of this associative nature we can correctly reduce the partial attention outputs corresponding to these unequally sized blocks.

\begin{figure*}
\includegraphics[width=\textwidth]{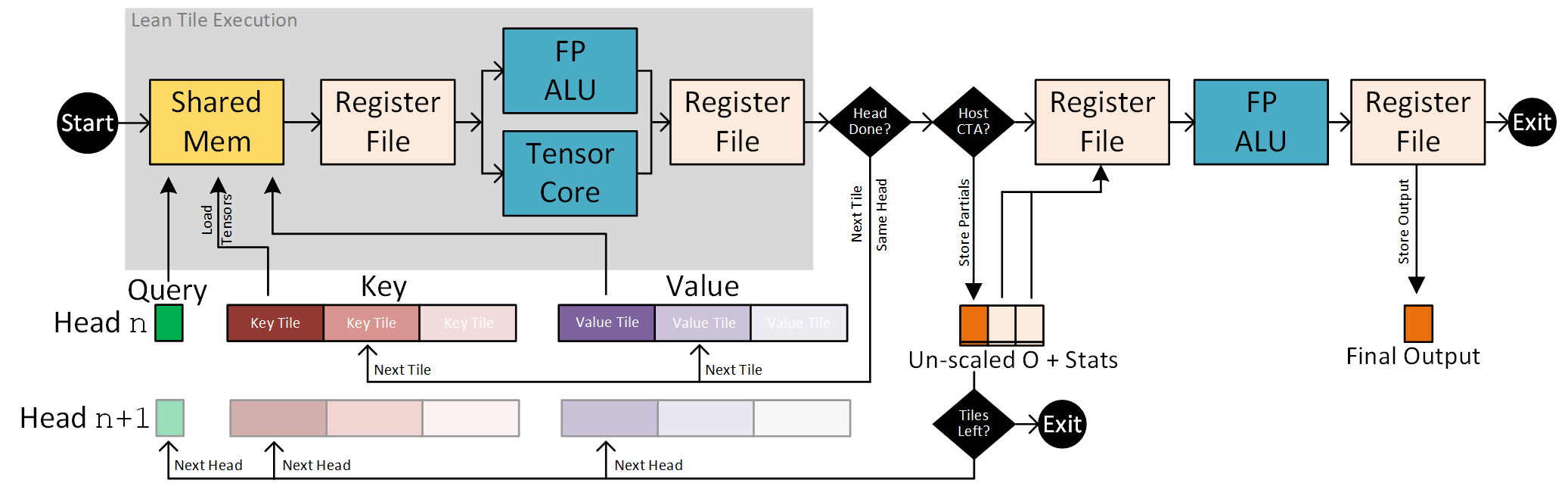}
\vspace{-0.2in}
\caption{Control and dataflow of a single CTA in LeanAttention utilizing various hardware resources. The tensors are loaded to shared memory in a tiled manner. At the end of a head, a reduction is performed if it is a host CTA or the partial un-scaled results are written to memory before moving to the next head.}
\vspace{-0.2in}

\label{fig:dataflow}
\end{figure*}

In the following subsections, we first outline the identification of softmax re-scaling as a reduction operation, followed by a conceptualization of a \textit{LeanTile} as a unit granularity in a CTA block and the stream-K style mapping within these CTAs, followed by an explanation of the overall execution flow of LeanAttention. \autoref{fig:dataflow} shows the high-level architectural execution flow of a single CTA computing LeanAttention.

\subsection{Softmax Re-scaling as Reduction}
\label{subsec:softmaxreduction}

% FlashAttention-2 uses the online softmax technique~\cite{milakov2018online} to split the attention computation for a single query block into chunks of work. Each chunk of work comprises a key block and a corresponding value block, and these chunks arrive in a sequential manner to update the attention output for the given query block. 
% FlashAttention-2 computes online softmax for every incoming chunk, rescales the intermediate output obtained from the previous chunk and combines it with the partial output from the current chunk to get the latest updated output. However, this method of computing exact attention is constrained in its sequentiality, leading to slower computation especially in the decode phase where there are a lot of key/value chunks to parse through for a given query block. 

% Due to its sequential manner of updating an attention output tile, FlashAttention-2 suffers from slower computation during the decode phase where there are a lot of key/value blocks to parse through sequentially to compute the output tile corresponding to a given input query tile.

In LeanAttention, we propose computation of partial attention outputs of a given query tile concurrently on different hardware units, while ensuring that we have a well-balanced work distribution across all hardware units through a Stream-K style decomposition of attention (discussed later in \autoref{subsec:streamkmapping}). This decomposition results in splits of work for a given SM that are not always equal in size, i.e., the key/value tensors of a given query tile are not dispatched in same-sized blocks to different SMs (unlike FlashDecoding~\cite{flashdecoding}, FlashInfer~\cite{flashinfer}). For example, in \autoref{fig:mainfig}, for computing LeanAttention for a query tile $h_0$, SM0 and SM1 receive same-sized KV blocks (each KV block consists of 2 LeanTiles), but SM2 receives half the amount of work for $h_0$ (the KV block for it consists of 1 LeanTile) than SM0 or SM1 received.

To reduce these partial attention outputs that result from differently sized blocks, we use a softmax re-scaling operation. This requires us to identify softmax re-scaling's associativity property that allows it to correctly reduce blocks of unequal sizes, i.e., application of softmax re-scaling as a reduction operator will give the \emph{same exact attention output} with no loss in accuracy, \textit{regardless of the way the work might be split, whether in same-sized blocks or arbitrary differently sized blocks}.

Without loss of generality, we describe this process of reduction to obtain one row vector of the attention score matrix $\vS$, of the form $\begin{bmatrix} \vS^{(x)} & \vS^{(y)} \end{bmatrix}$ consisting of some unequal length vectors
$\vS^{(x)}, \vS^{(y)}$ where $\vS^{(x)} \in \mathbb{R}^{1 \times B_c^{(x)}}$ and $\vS^{(y)} \in \mathbb{R}^{1 \times B_c^{(y)}}$, where 1 is the query length and $B_c^{(x)}$ and $B_c^{(y)}$ are the unequal context lengths. The vectors $\vS^{(x)}$ and $\vS^{(y)}$ were computed from $\vQ \times (\vK^{(x)})^{T}$ and $\vQ \times (\vK^{(y)})^{T}$ as shown in \autoref{fig:leanmath}. Note that, to generalize this procedure for blocks of any size, the context length of $ \vK^{(x)}$ and $\vK^{(y)}$ are $B_c^{(x)}$ and $B_c^{(y)}$ and are not necessarily equal.

The attention computation is split into two parts. The first part involves calculation of an ``un-scaled" version of $\vO^{(i)}$ (where $i$ is either $x$ or $y$) along with statistics $m^{(i)}$ and $\ell^{(i)}$: 

\vspace{-0.2in}
\begin{align*}
 & \vS^{(i)} = \vQ (\vK^{(i)})^{T} \in \mathbb{R}^{1 \times B_c^{(i)}}\\
& m^{(i)} = \mathrm{rowmax}(\vS^{(i)})  \in \mathbb{R}^{1 \times 1}\\
 & \ell^{(i)} = \mathrm{rowsum}(e^{\vS^{(i)} - m^{(i)}}) \in \mathbb{R}^{1 \times 1} \\
 & A^{(i)} = exp(\vS^{(i)} - m^{(i)}) \in \mathbb{R}^{1 \times B_c^{(i)}}\\
 & {\tilde{\vO}}^{(i)} = A^{(i)} \vV^{(i)} \in \mathbb{R}^{1 \times d}\\
\end{align*}
\vspace{-0.2in}

\noindent{\textbf{Softmax Re-scaling Operation.}} 
The second part involves re-scaling the ``un-scaled" outputs ${\vO}^{(\mathrm{i})}$ using the previously computed statistics $m^{(i)}$ and $\ell^{(i)}$. \\
We define the softmax re-scaling operation $f(x,y)$ for two intermediate outputs $\vO^{(x)}$ and $\vO^{(y)}$ as follows: \\

\vspace{-0.2in}
\begin{align*}
  & m^{(x,y)} = \max(m^{(x)}, m^{(y)}) \\
  & \ell^{(x,y)} = e^{m^{(x)} - m^{(x,y)}} \ell^{(x)} + e^{m^{(y)} - m^{(x,y)}} \ell^{(y)} \\
  & f(x,y) = \diag(e^{m^{(x)} - m^{(x,y)}}){\tilde{\vO}}^{(x)} + \diag(e^{m^{(y)} - m^{(x,y)}}){\tilde{\vO}}^{(y)} \\
  & f(x,y) = {\tilde{\vO}}^{(x,y)} \\
  & \vO^{(x,y)} = \diag(\ell^{(x,y)})^{-1}{f(x,y)} \\
\end{align*}
\vspace{-0.2in}

\noindent{\textbf{Proof of Associativity}} 
The associative nature of softmax re-scaling $f(x,y)$ allows us to reduce intermediate outputs produced from key/value vectors of different lengths in LeanAttention. We shall briefly prove that $f(f(x,y), z) = f(x, f(y,z)) = f(x,y,z)$, where:
$f(x,y) = {\tilde{\vO}}^{(x,y)}$, $f(y,z) = {\tilde{\vO}}^{(y,z)}$ and $f(x,y,z) = {\tilde{\vO}}^{(x,y,z)}$. 
% The key vectors $K^{(x)}$, $K^{(y)}$ and $K^{(z)}$ and value vectors $V^{(x)}$, $V^{(y)}$ and $V^{(z)}$ are of lengths $B_c^{(x)}$, $B_c^{(y)}$ and $B_c^{(z)}$ that are not necessarily equal.
 \\

Proving that $f(f(x,y), z) = f(x,y,z)$:
\begingroup
\allowdisplaybreaks
\begin{align*}
 f(x,y) &= {\tilde{\vO}}^{(x,y)} \\
f(f(x,y), z) &= \diag(e^{m^{(x,y)} - m^{((x,y),z)}}){\tilde{\vO}}^{(x,y)} \\
&+ \diag(e^{m^{(z)} - m^{((x,y),z)}}){\tilde{\vO}}^{(z)} \\
&= \diag(e^{m^{(x,y)} - m^{(x,y,z)}}){\tilde{\vO}}^{(x,y)} \\
&+ \diag(e^{m^{(z)} - m^{(x,y,z)}}){\tilde{\vO}}^{(z)} \\
&= \diag(e^{m^{(x,y)} - m^{(x,y,z)}}) \\
& \times (\diag(e^{m^{(x)} - m^{(x,y)}}){\tilde{\vO}}^{(x)} \\
& + \diag(e^{m^{(y)} - m^{(x,y)}}){\tilde{\vO}}^{(y)}) \\
& + \diag(e^{m^{(z)} - m^{(x,y,z)}}){\tilde{\vO}}^{(z)} \\
&= \diag(e^{m^{(x)} - m^{(x,y,z)}}){\tilde{\vO}}^{(x)} \\
& + \diag(e^{m^{(y)} - m^{(x,y,z)}}){\tilde{\vO}}^{(y)} \\
& + \diag(e^{m^{(z)} - m^{(x,y,z)}}){\tilde{\vO}}^{(z)} \\
&= {\tilde{\vO}}^{(x,y,z)} = f(x,y,z)\\
\end{align*}
\endgroup

Therefore, $f(f(x,y), z) = f(x,y,z)$ and similarly $\ell^{((x,y),z)} = \ell^{(x,y,z)}$. For brevity, we omit the proof of $f(x, f(y, z)) = f(x,y,z)$, but it can deduced in a similar manner.

This associativity of softmax re-scaling is leveraged in LeanAttention to concurrently calculate the ``partial" outputs produced from unequally sized KV blocks and then ``reduce" them to obtain exact attention. \\

\subsection{LeanTile}
\label{subsec:leantile}
To enable us to efficiently distribute the work of computing the attention output tiles, we define the smallest granularity of a KV block as a \textit{LeanTile}. A single LeanTile iteration computes ``local attention" across a subset of tokens along the $N_k$ dimension as shown in the grey box of \autoref{fig:dataflow}. Thus, a LeanTile takes in a query, key, and value tensor and computes local attention to generate an un-scaled attention output.
% A CTA in Lean attention constitutes many such LeanTiles terations belonging to either a single output tile or many different output tiles. 

% A partial output tile computed by a single CTA is a sequential updation and accumulation of many lean tile iterations present within the boundaries of the partial output tile. This essentially behaves as FlashAttention-2 for that partial output tile, assuming the output tile boundaries to be the output matrix boundaries in algorithm~\ref{alg:flashattn}. 

% The Stream-K partitioning in Lean Attention may result in CTA's spanning subsets of inner mode iterations for more than one output tile. Each of this inner mode iterations occurs at a granularity that we define as a \textit{Lean Tile}. A CTA block in lean attention constitutes many such lean tile iterations belonging to either a single output tile or many different output tiles. 

% A single lean tile iteration computes ``local attention" for that iteration. A partial output tile computed by a single CTA is a sequential updation and accumulation of many lean tile iterations present within the boundaries of the partial output tile. This essentially behaves as FlashAttention-2 for that partial output tile, assuming the output tile boundaries to be the output matrix boundaries in algorithm~\ref{alg:flashattn}. 

Algorithm~\ref{alg:leantile} depicts the subroutine for computing the partial attention outputs for a sequence of LeanTile's. This LeanTile() subroutine is called when computing each partial output tile in a CTA launched in LeanAttention, as will be discussed later (Algorithm~\ref{alg:leanattnalgstreamK}). 

% LeanTile Algorithm
\begin{algorithm}[tb]
   \caption{LeanTile() for a sequence of lean tile iterations.}
   \label{alg:leantile}
\begin{algorithmic}[1]
\STATE \textbf{function} LeanTile(tile\_idx, iter\_begin, iter\_end)
\STATE \textbf{\_shared\_} $O_{acc}[T_m, d]$
\STATE \textbf{\_shared\_} $Q_{f}[T_m, d]$
\STATE \textbf{\_shared\_} $K_{f}[T_n, d]$
\STATE \textbf{\_shared\_} $V_{f}[T_n, d]$
\STATE \textbf{\_shared\_} $m[T_m, 1]$
\STATE \textbf{\_shared\_} $l[T_m, 1]$
% \STATE To find output tile O's coordinates:
\STATE Initialize $O_{acc}$ to $(0)_{T_m \times d}$ $\in R^{T_m \times d}$ in SMEM.
\STATE Initialize $m$ to $(-\infty)_{T_m \times 1}$ and $l$ to $(0)_{T_m \times 1}$ $\in R^{T_m \times 1}$ in SMEM.
\STATE $mm = T_m \times$ (tile\_idx / 1) 
% (since O isn't partitioned along headdim). %\RS{check this once}
\STATE $nn = d \times$ (tile\_idx \% 1)
\STATE Perform lean tile iterations for this output tile.
\FOR{$iter = iter\_begin$ {\bfseries to} $iter\_end$}
\STATE $kk = iter \times T_n$
\STATE load fragments from GMEM to SMEM
\STATE $Q_{f} = LoadFragment(Q, mm, nn)$
\STATE $K_{f} = LoadFragment(K, nn, kk)$
\STATE $V_{f} = LoadFragment(V, nn, kk)$
\STATE Compute on chip: 
\STATE $S_{f} = Q_{f} K_{f}$ where $S_{f} \in R^{T_m \times T_n}$
\STATE $m^{new} = max(m, rowmax(S_{f}))$
\STATE $P_{f} = exp(S_{f} - m^{new})$  where $P_{f} \in R^{T_m \times T_n}$
\STATE $l^{new} = e^{m - m^{new}}l + rowsum(P_{f})$
\STATE $O_{acc} = P_{f} V_{f} + diag(e^{m - m^{new}}) O_{acc}$
\STATE $l = l^{new}, m = m^{new}$
\ENDFOR
\STATE \textbf{return} $O_{acc}$, $l$, $m$
\STATE \textbf{end function}
\end{algorithmic}
\end{algorithm}

To efficiently split attention into smaller tiles, it is necessary to identify the smallest tile size capable of achieving the highest compute efficiency. LeanTile size depends on the computational power and memory access costs and, thus, are fixed for a particular hardware architecture. After an extensive empirical sweep through various sizes for a \textit{LeanTile}, we found a tile size granularity of 256 and 128 tokens along the $N_k$ dimension to be the most optimal for a head size of 64 and 128 respectively for FP16\textrightarrow32 problems while experimenting on an A100 GPU \cite{nvidiaA100, jia2021dissecting}. This optimal size can similarly be identified for other head dimensions and hardware architectures.

% Another point to note is that the stream-K partitioning scheme has been defined for general matrix multiplications (GEMMs), but since attention computation involves two batched matrix multiplications (BMMs), we concatenate batch instances and heads along the inner modes of each input matrix, resulting in one large MatMul. \RS{Is this the right way to phrase this?}

\subsection{Decomposition and Mapping of LeanTiles}
\label{subsec:streamkmapping}
Finally, LeanAttention uses a stream-K~\cite{streamk} style decomposition and mapping of these LeanTiles to deliver efficient execution of attention. \\

\noindent{\textbf{Stream-K Decomposition}}. Stream-K is a parallel decomposition technique for dense matrix-matrix multiplication on GPUs. Stream-k partitioning addresses the inefficiencies in fixed-split by dividing the total workload (MAC operations) equally to all the CTAs using a pre-determined optimal tile size. It does this by rolling out the inner mode iterations of all output tiles and appending them along the inner mode to form a linear mapping. With the given grid size, it divides this total work into buckets demarcated appropriately such that each CTA has equal amount of MAC operations to perform. The grid size is fixed for a given tile size as LeanAttention provides equal work to all SMs. For example, for a tile of 256 tokens, two CTAs can be co-executed in a single wave with the available shared memory of A100 GPU. This would result in a total grid size of $108 (NumSMs) \times 2 = 216$. Number of tiles to be computed by each CTA can, thus be calculated as follows:
\begin{equation}
\footnotesize
TilesPerCTA = \frac{BatchSize \times NumHeads \times ContextLength}{TileSize \times NumSMs \times MaxCTAsPerSM}
\end{equation}

% This grid size is determined by heuristics that sweep through all possible grid sizes and finds the most optimal one which enables extensive parallelism and optimal wave quantization that compensate well for any overhead that comes from reduction of the partial outputs.
%Since the grid size is fixed, it implies the existence of cases where a given CTA's boundaries might not coincide with the iteration boundaries for a given output tile. In such cases, the results of the partial output tile computed are to be accumulated or ``fixed-up" with the other portions of the same output tile computed by other CTAs. A given CTA could be in charge of computing a subset of iterations along K-mode for one or more independent output tiles, and the number of iterations of each partial tile it computes need not be equal, but the sum of them must be equal to the work distributed to each CTA. 

LeanAttention extends Stream-K style of linear mapping of iterations by rolling out LeanTile iterations in a similar fashion, assigning equal number of LeanTiles to every CTA as shown in \autoref{fig:mainfig}. Each CTAs range of LeanTile iterations is mapped contiguously into the batch size \textrightarrow heads \textrightarrow context length linearization, crossing the head and query boundary as it may. Should a given CTA's starting and/or ending LeanTile not coincide with the head's boundary, it must consolidate its partial output with those of the other CTA(s) also covering that head's output tile. In our implementation of LeanAttention, each attention output tile is consolidated by the CTA that performed that output's first LeanTile (called as a host block). Before it can do so, however, it must accumulate the un-scaled output tensors from other CTA(s) in temporary global storage, as shown in \autoref{fig:mainfig}. The negligible synchronization overhead of original stream-K implementation also extends to LeanAttention, thus leading to near 100\% occupancy of SMs (not tensor core utilization) during the execution of a single CTA. Note that the temporary global storage overhead is minimal in the case of decode-phase where the output tensors are of dimensions $1 \times head\_dim$, where $head\_dim$ is typically in the range of 64 to 256.

Further, since we distribute the overall attention problem into optimal LeanTiles, we achieve a near 100\% quantization efficiency irrespective of problem size (context length). This cohesive implementation of parallel computation and reduction happens in a single kernel launch in LeanAttention, avoiding the reduction kernel launch overheads that FlashDecoding suffers from. A difference in Stream-K decomposition in LeanAttention is in the reduction or ``fix-up" phase. While Stream-K for MatMuls has addition as its reductive operator, LeanAttention has softmax re-scaling as its reductive operator.

Naturally, some CTA's will be computing LeanTile iterations of more than one independent output tile. In such cases, stream-K's equalized partitioning makes lean attention more adept for problem sizes which would not occupy the hardware well if executed using its counterparts, FlashAttention-2 and FlashDecoding. To enable such a smooth transition between tiles, the input tensor view is also different in LeanAttention compared to FlashAttention-2. This requires a constant stride moving between different heads as we transition from the LeanTile of one head to another requiring query, key, and value tensors be of the shape $(batch\_size, heads, query/ctx\_length, head\_dim)$ compared to FlashAttention-2's requirement of $(batch\_size, query/ctx\_length, heads, head\_dim)$.

With this design of execution, we must point out that LeanAttention behaves as a versatile attention partitioning mechanism which generalizes to FlashAttention-2 in the case where the number of output tiles is equal to grid size, and generalizes to FlashDecoding when grid size is an even multiple of number of output tiles. Finally, for all other cases (most common) LeanAttention efficiently distributes the work across the compute resources available in the system. Thus, LeanAttention will either always perform better or the same as FlashAttention-2 and FlashDecoding.

\begin{figure}[t]
\begin{center}
\centerline{\includegraphics[width=\columnwidth]{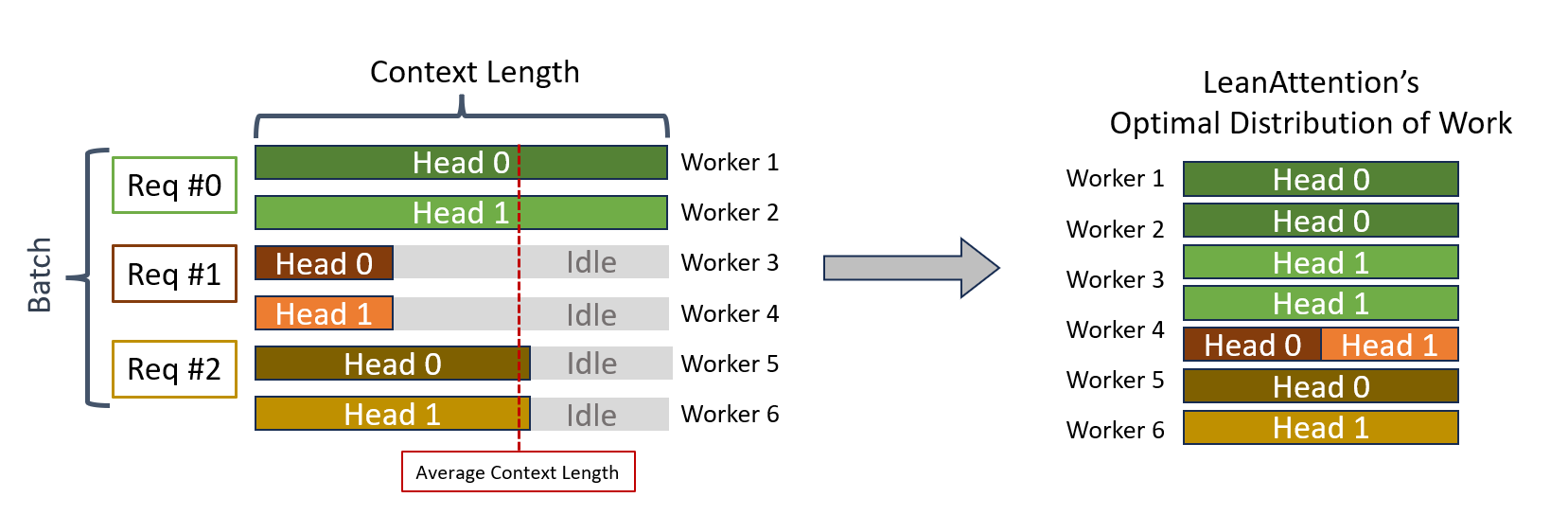}}
\vspace{-0.2in}
\caption{Illustrative diagram of LeanAttention's optimal distribution of work in the ragged batching case. Each SM receives equal amount of LeanTiles.}
\label{fig:raggeddist}
\end{center}
\vspace{-0.3in}
\end{figure}

\noindent{\textbf{Lean Ragged Batching}}. For the special case of dealing with unequal context lengths within a batch of requests, the number of LeanTiles per request becomes unique, resulting in a total workload that is smaller compared to the non-ragged case. To account for this difference, ragged key/value tensor inputs to LeanAttention are first prepared with unpadded dimensions of $(NumHeads, TotalContextLength, HeadDim)$, where $TotalContextLength$ is the sum of all distinct context lengths within the batch. As seen, the batch dimension is eliminated, and both batch indices and the true context lengths of each request are tracked through pointers to a cumulative sequence lengths array for each input tensor. These pointers have size $(BatchSize + 1)$ each, introducing minimal memory overhead.

With the total workload of LeanTiles correctly determined, Lean ragged batching functions identically to the non-ragged case as shown in . The workload is distributed evenly across the grid as shown in \autoref{fig:raggeddist}, ensuring each CTA receives the same number of LeanTiles to process. The range of LeanTile iterations assigned to each CTA is mapped contiguously in a Heads $\rightarrow$ TotalContextLength linearization and partial outputs for each head are consolidated in the same manner as in the non-ragged case.

\subsection{Execution Flow}
\label{subsec:executionflow}
Algorithm~\ref{alg:leanattnalgstreamK} depicts a Stream-K style execution of LeanAttention. For a fixed grid size $G$, CTAs are launched and given equal amount of LeanTile iterations to work with (Line 7). Each CTA computes the LeanTile() subroutine for every distinct output tile that comes under its boundaries (Line 16). \autoref{fig:dataflow} shows the execution flow of a single CTA computing partial and final attention outputs for the assigned heads.

The unique reduction phase of LeanAttention, characterized by its softmax re-scaling operator, is performed by the host CTA block (Lines 24-40). A host CTA (Line 17) is the CTA responsible for computing the first ever LeanTile for a given output tile, and it behaves as the reducing CTA during parallel reduction of partial un-scaled outputs. 

All non-host CTAs will share their partials through a store to global memory and signal their arrival (Lines 20-23). On the other hand, a host block, which is a non-finishing block (Lines 24-25), needs to wait for other contributing peer CTA blocks to signal their completion (Line 28) and then proceed to carry out the reduction (Lines 29-35). 

A CTA that is computing the attention for a head exclusively completes all the LeanTile iterations for its output tile in a single CTA and so can directly store its results from the register file to global memory (Line 38-39) without any need for further reduction. 

% LeanAttention Algorithm
\begin{algorithm}[tb]
   \caption{Lean Attention}
   \label{alg:leanattnalgstreamK}
\begin{algorithmic}[1] 
\STATE \textbf{\_shared\_} $O[T_m, d]$
\STATE \textbf{\_shared\_} $m[T_m, 1]$
\STATE \textbf{\_shared\_} $l[T_m, 1]$
\STATE Number of output tiles: $C_m = \lceil N_q / T_m \rceil$
\STATE Number of iterations for each output tile: $C_n = \lceil N_k / T_n \rceil$
\STATE Total number of iterations: $I = C_m C_n$
\STATE Number of iterations per CTA: $I_G = I / G$
% \FOR{$g=1$ {\bfseries to} $G$}
\STATE \textbf{fork} CTA$_g$ in $G$ \textbf{do}
\STATE cta\_start = $g$ I$_G$ and cta\_end = cta\_start + I$_G$
\FOR{iter = cta\_start {\bfseries to} cta\_end}
\STATE Index of current output tile: tile\_idx = iter / $C_n$
\STATE tile\_iter = tile\_idx $\times C_n$ 
\STATE tile\_iter\_end = tile\_iter + $C_n$
\STATE local\_iter = iter - tile\_iter
\STATE local\_iter\_end = min(tile\_iter\_end, cta\_end) - tile\_iter
\STATE O, m, l = LeanTile(tile\_idx, local\_iter, local\_iter\_end)
% \STATE Reduction of partial tiles across CTAs:
\STATE host-block if: iter == tile\_iter
\STATE finishing-block if: cta\_end $>=$ tile\_iter\_end
\IF{!(host-block)}
\STATE StorePartials(Op[g], O)
\STATE StorePartials(mp[g], m)
\STATE StorePartials(lp[g], l)
\STATE Signal(flags[g])
\ELSE
\IF{!(finishing-block)}
\STATE last\_cta = tile\_iter\_end / $C_n$
\FOR{cta = (g + 1) {\bfseries to} last\_cta}
\STATE Wait(flags[cta])
\STATE $m_{cta}$ = LoadPartials(mp[cta])
\STATE $l_{cta}$ = LoadPartials(lp[cta])
\STATE $O_{cta}$ = LoadPartials(Op[cta])
\STATE $m^{new} = max(m_{cta}, m)$
\STATE $l^{new}$ = $e^{m_{cta} - m^{new}}l_{cta} + e^{m - m^{new}}l$
\STATE $O^{new}$ = $e^{m_{cta} - m_i^{new}}O_{cta} + e^{m - m_i^{new}}O$
\STATE Update $m = m_i^{new}, l = l_i^{new}$
\ENDFOR
\ENDIF
\STATE Write $O = diag(l)^{-1}O$ to GMEM.
\STATE Write $L = m + log(l)$ to GMEM.
\ENDIF
\STATE iter = tile\_iter\_end
\ENDFOR 
\STATE \textbf{join}
\end{algorithmic}
\end{algorithm}

\begin{figure*}
\includegraphics[width=\textwidth]{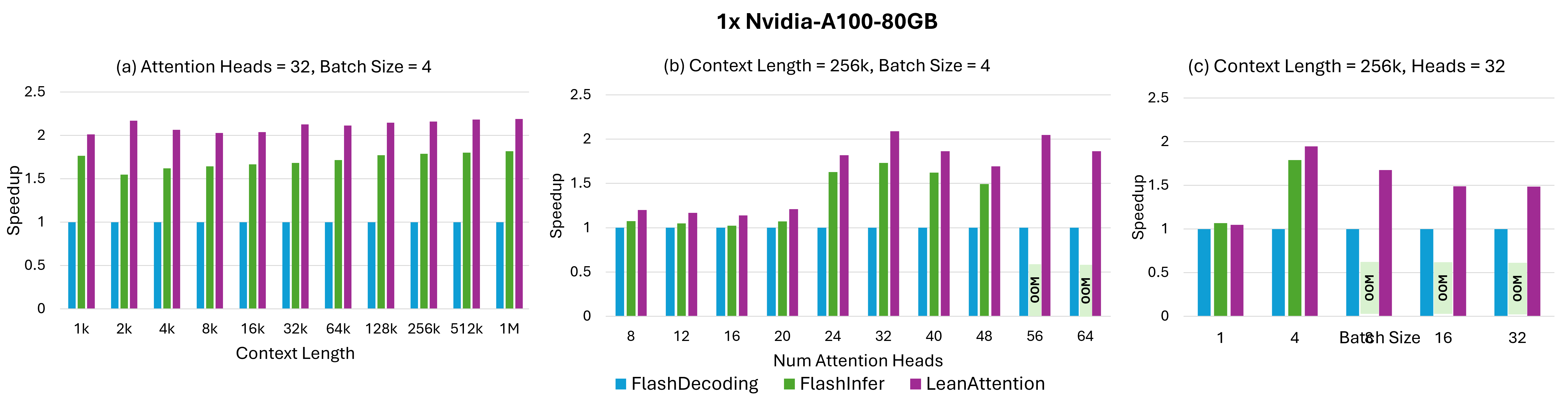}
\vspace{-0.3in}
\caption{Speedup of LA compared to state-of-the-art Attention execution mechanisms at different context lengths, batch sizes and attention heads with head dimension = 64 on a single Nvidia-A100-80GB GPU.}
\vspace{-0.1in}
\label{fig:1xUnitResults}
\end{figure*}

\section{Evaluation Methodology}
\label{sec:evalmethod}
% \section{Compute Cost Evaluation}
% \label{computecost}
\noindent{\textbf{Implementation.} We implement LeanAttention using the CUTE abstractions~\cite{cutelayout, cutetensor, cutemma} provided by Nvidia's CUTLASS library~\cite{cutlassgithub}. For comparative measurements we utilize FlashAttention-2's implementation of FlashDecoding as it is available on their Github repository~\cite{githubGitHubDaoAILabflashattention}\footnote{Version 2.5.6} and FlashInfer's implementation from their Github repository~\cite{flashinfer}\footnote{Version 0.1.6 - Note that we increase the number of heads in the single batch decoding API to simulate batch size as the batch API performance is too low}. For the end-to-end inference results we use OPT models as available in the HuggingFace Transfomers repository~\cite{huggingface} and modify them to allow execution via LeanAttention wherever necessary. Note that optimizations such as FlashAttention-3~\cite{flashattention3} are orthogonal to this work and targeted specifically for H100. The core computation of LeanAttention can adopt FlashAttention-3's optimizations for further benefits on the Nvidia-H100 GPU~\cite{luo2024benchmarkingdissectingnvidiahopper}.

\noindent{\textbf{System.} We benchmark the attention mechanisms on Nvidia-A100-80GB-GPU~\cite{nvidiaA100} system with up to 8 GPUs. We measure runtime using a single GPU as well as 8xGPUs for larger models and context lengths. A single A100 GPU consists of 108 streaming multiprocessors (SMs) with an 80GB HBM global memory. To demonstrate LeanAttention's versatility across hardware architectures we benchmark it similarly on a single Nvidia-H100-SXM-80GB-GPU~\cite{luo2024benchmarkingdissectingnvidiahopper} which has 132 streaming multiprocessors and an 80GB HBM global memory.

% NVIDIA's open source CUTLASS library provides templated abstractions for blocking GEMMs and clean layouts through CuTe templates for accessing these blocked partitions. We've exploited these utilities heavily in implementing Lean Attention. 
\noindent{\textbf{Batching.}} The evaluations assume the same context length for all queries in a batch working in tandem with batching techniques such as Orca\cite{orca22}. However, LeanAttention supports varied context length execution including heterogeneous batching such as prefill queries with decode.

\noindent{\textbf{Multi-GPU Tensor Parallelism.}} We utilize Tensor Parallelism for the multi-GPU measurements to reflect the large model executions. Since FlashDecoding does not support Tensor Parallelism, we scale the implementation to the total number of SMs available in the system.

\noindent{\textbf{Attention Mechanism.} In addition to FlashDecoding~\cite{flashdecoding}, we also benchmark FlashInfer for comparison against LeanAttention. FlashInfer has two kernel implementations for decoding: one for single-request decoding and another for batched-request decoding. The single-request decode kernel implements Fixed-Split, while the batched decode kernel implements PagedAttention. In our experiments, we benchmarked the batched decode kernel using a page size of 16. We've observed no impact of page size on FlashInfer’s latency, a finding consistent with their blog post~\cite{flashinfer}. Moreover, since FI reserves extra GPU memory to store buffers for managing its paged KV cache, we were unable to test it on certain large problem sizes, which we indicate as Out-of-Memory (``OOM") errors in our evaluation figures. For the rest of the paper we refer to FlashDecoding as FD, FlashInfer as FI and LeanAttention as LA.

\begin{figure}
\begin{center}

\includegraphics[width=0.95\columnwidth]{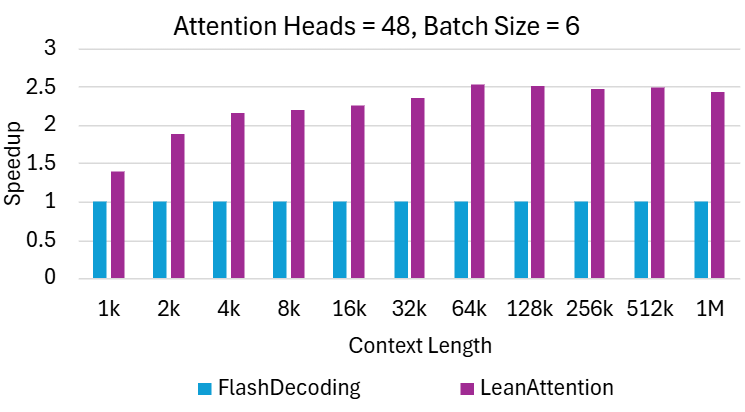}
\vspace{-0.2in}
\caption{Speedup of LA compared to state-of-the-art Attention execution mechanisms at varying context lengths, at a fixed batch size and attention heads with head dimension = 64 on a single Nvidia-H100-SXM-80GB GPU.}
\vspace{-0.2in}
\label{fig:1xUnitResultsH100}
\end{center}
\end{figure}

\section{Evaluation Results}
\label{sec:evalresults}
In this section, we evaluate the impact of LeanAttention (LA) at the attention operation-level as well as end-to-end inference performance.

\subsection{Benchmarking Attention - Decode-Phase}
\label{subsec:attenperf}

We benchmark the runtime of just the attention operation using the different mechanisms at varying context lengths, number of attention heads, head dimensions (64:default and 128), and inference batch sizes on a single Nvidia A100-80GB GPU and a single Nvidia H100-SXM-80GB GPU.

\noindent{\textbf{Increasing Context Length.} \autoref{fig:1xUnitResults}(a) shows the speedup of different attention mechanisms for a model with 32 attention heads with a  batch size of 4 on a single A100 GPU. LA delivers close to 2x speedup compared to FD even at smaller context lengths, reaching up to 2.18x speedup as the context lengths grows to 256k tokens. When context lengths exceed 16k, we observe more than 1.46x speedup over FI. Repeating a similar exercise on an H100 GPU, we observe the speedups of FD versus LA at a fixed batch size of 6 and 48 attention heads as shown in \autoref{fig:1xUnitResultsH100}(a). LA delivers more than 2x speedup even for contexts over 4k tokens, reaching upto a maximum of 2.52x speedup over FD at a 64k context length and 4.48x speedup over FI which more or less plateaus at the context lengths increase.

\begin{figure*}[t]
\includegraphics[width=\textwidth]{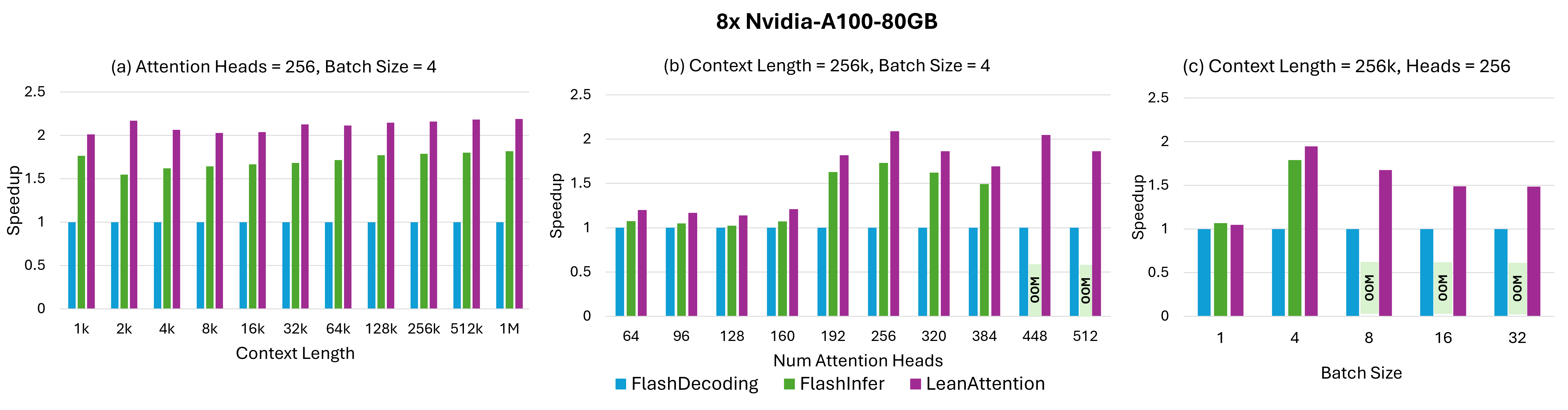}
\vspace{-0.3in}
\caption{Speedup of LA compared to state-of-the-art Attention execution mechanisms at different context lengths, batch sizes, attention heads with head dimension = 64} on an 8x Nvidia-A100-80GB system.

\label{fig:8xUnitResults}
\vspace{-0.3in}
\end{figure*}

\begin{figure}
\begin{center}

\includegraphics[width=0.95\columnwidth]{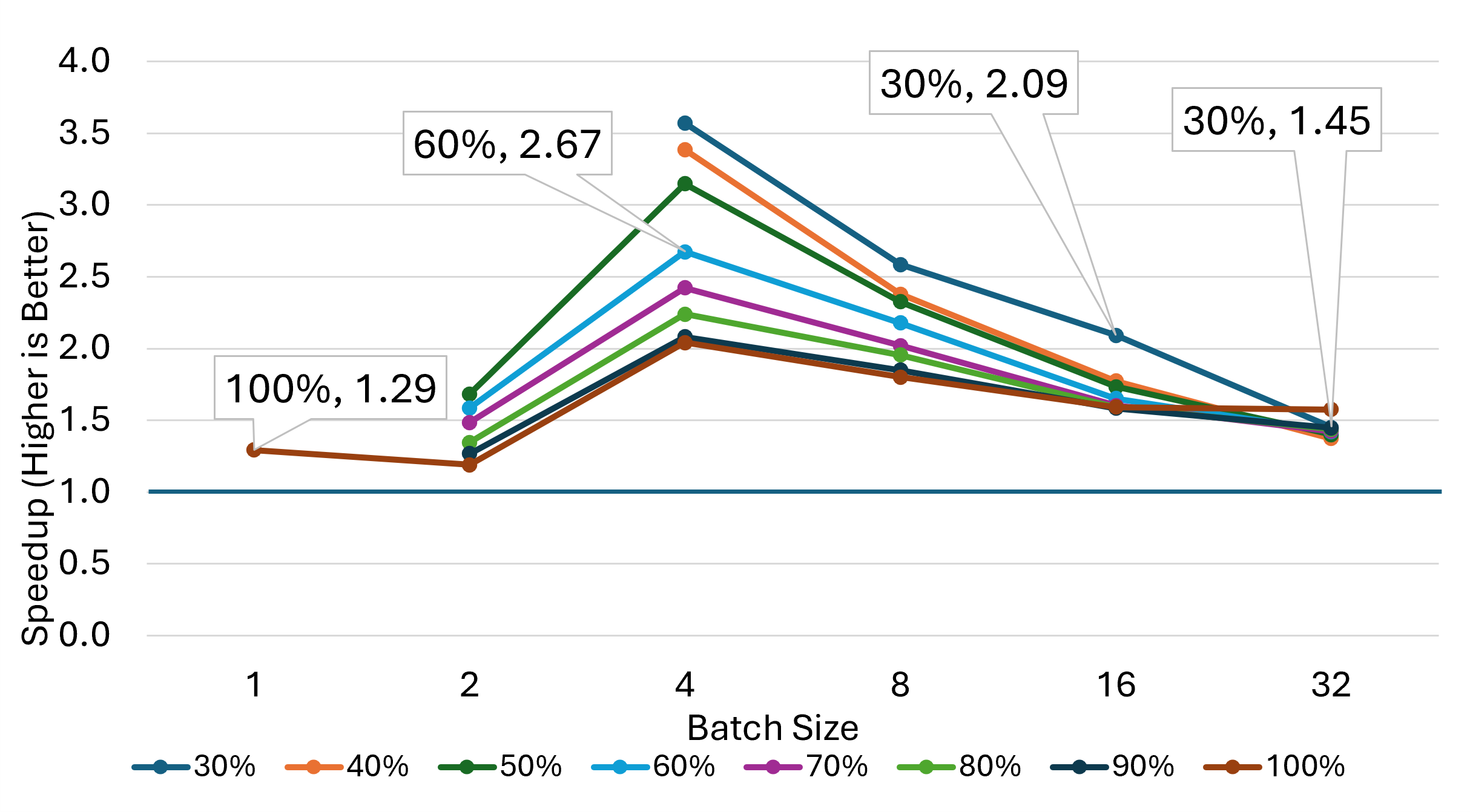}

\caption{Speedup offered by LA over FD at different batch sizes with heterogeneous context lengths. Batch context ratio(\%) shows the ratio of average context length over maximum context length }
\vspace{-0.2in}
\label{fig:raggedbatch}
\end{center}
\end{figure}

\begin{figure}
\begin{center}

\includegraphics[width=0.95\columnwidth]{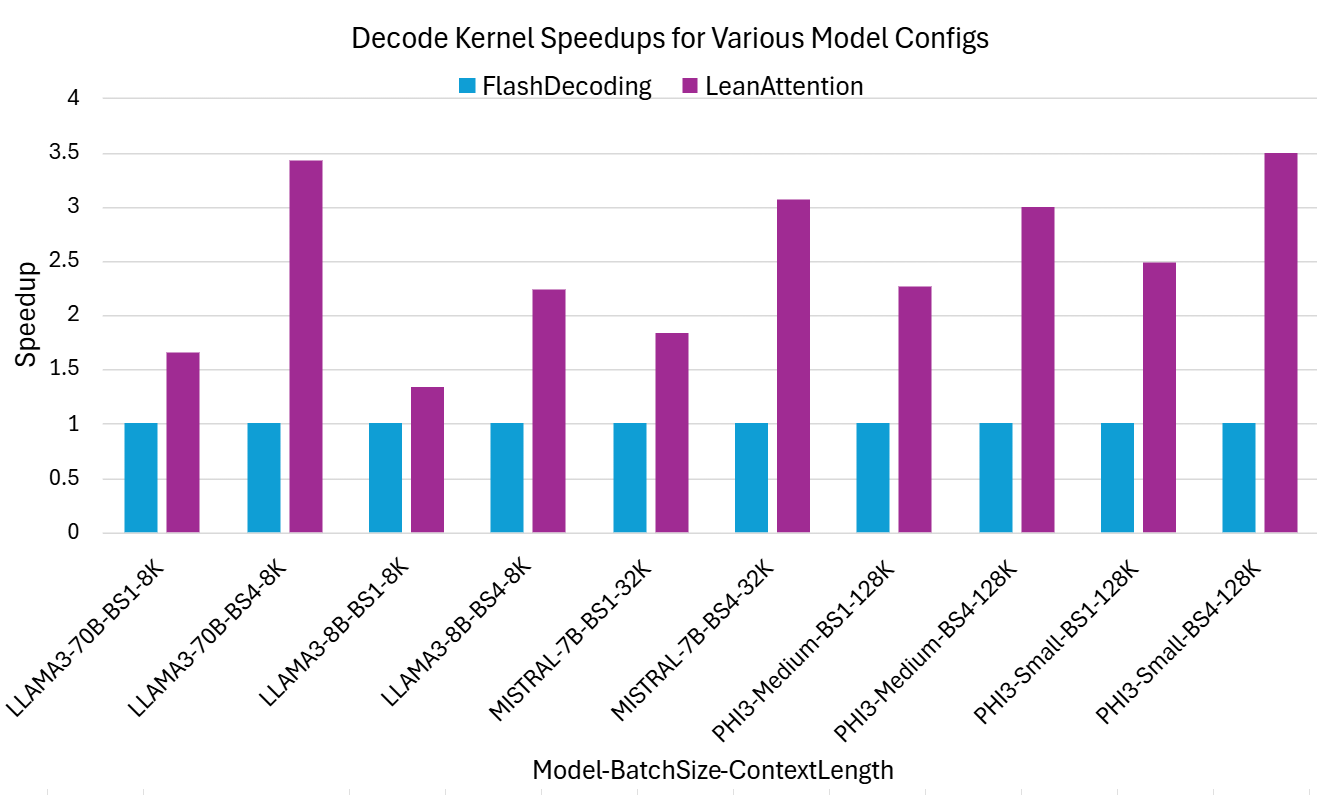}

\caption{Speedup offered by LA for decode attention across models, batch sizes, context lengths integrated with ONNXRT}
\vspace{-0.3in}
\label{fig:headdim128}
\end{center}
\end{figure}

\begin{figure}[t]
\begin{center}
\includegraphics[width=0.9\columnwidth]{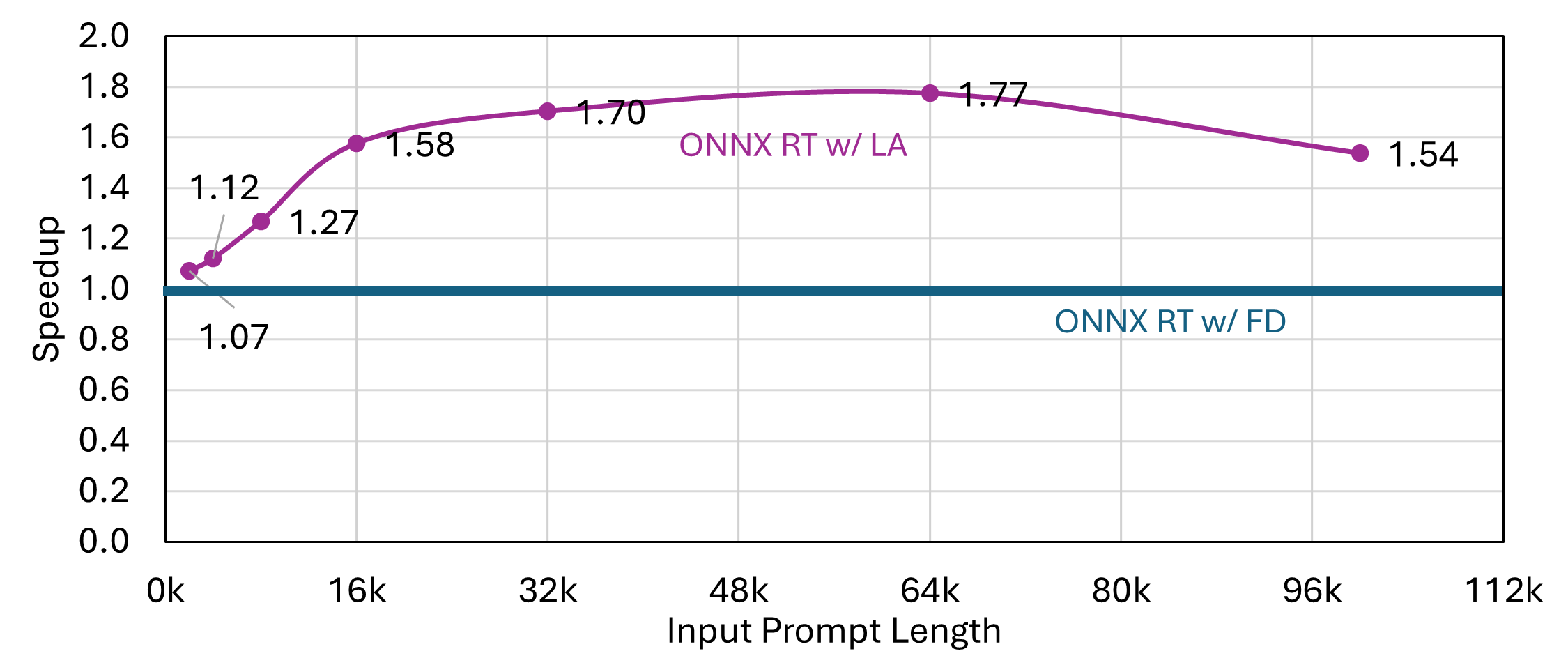}
\vspace{-0.1in}
\caption{End-to-End Speedup of LA compared to FD in ONNXRT running Phi-3 Medium model at different context lengths, batch size = 1, prompt size : output tokens = 8 : 1}
\vspace{-0.3in}
\label{fig:EndToEndSpeedup}
\end{center}
\end{figure}

\noindent{\textbf{Increasing Attention Heads.}} \autoref{fig:1xUnitResults}(b) and \autoref{fig:1xUnitResultsH100}(b) shows the speedup delivered by LA compared to FD and FI for models with an increasing number of heads. LA delivers comparable speedups to FD and FI at smaller model sizes. With fewer attention heads, FD’s fixed-split mechanism can distribute the workload as evenly as LA. However, as the number of heads increases, FD resorts to fewer splits per head, resulting in partially filled waves of attention on the SMs. In contrast, LA maintains even workload distribution at both small and large model sizes. The speedups over FD and FI vary depending on model size as depicted, with LA achieving an impressive 2.53x speedup over FD on an H100 GPU with 6 batched contexts of length 64k. Additionally, LA delivers more than 2x speedup over FD on an A100 GPU when the number of attention heads exceeds 24 for 4 batched contexts, each with a length of 256k. This shows that LA is able to scale well for both a small and large number of heads.

\noindent{\textbf{Effect of Batching.}} \autoref{fig:1xUnitResults}(c) and \autoref{fig:1xUnitResultsH100}(c) shows the performance improvement of LA at varying batch sizes. As expected, we observe that LA gives comparable speedup to FD and FI. This is because both FD and FI are able to employ a higher number of splits at smaller batch sizes to occupy all the SMs in the GPU. However, as batch sizes increase, FD selects fewer splits per head. For instance, FD opts not to split at batch sizes above 4 in \autoref{fig:1xUnitResults}(c) and \autoref{fig:1xUnitResultsH100}(c) because the total number of heads in the batch exceeds the number of SMs available in the system. As a result, it behaves like vanilla FlashAttention-2, missing out on potential performance gains by leaving some SMs idle in its final, partially full wave. Consequently, LA achieves more than 1.5x speedup compared to FD through its stream-K-ed decomposition. 

Overall, we benchmarked the system on more than a 1,000 samples with varying batch sizes, context lengths, and attention heads. On an A100 GPU, we observed an average speedup of 1.73x over FD (Max: 2.18x for 56 heads, batch size 2, 256k context) and an average of 3.42x over FI (Max: 5.66x for 12 heads, batch size 8, 512k context). On the H100 GPU, we recorded an average speedup of 1.52x over FD (Max: 2.53x for 48 heads, batch size 6, 64k context) and an average speedup of 3.63x over FI (Max: 4.59x for 56 heads, batch size 4, 128k context). 

\noindent{\textbf{Ragged Batching in Decode.}}
For the purpose of our evaluations, we quantify the heterogeneity of a ragged batch as the ratio of average context length to the maximum context length present in the batch. \autoref{fig:raggedbatch} shows the speedup of LA over FD. We observe that as the heterogeneity of batch increases, LA delivers a higher speedup because of better distribution of work across SMs.

\noindent{\textbf{Multi-GPU Execution}} Repeating a similar benchmarking process on an 8xA100 GPU system, we vary the context lengths from 1k to 1M, with 256 attention heads at a batch size of 4 as shown in \autoref{fig:8xUnitResults}(a). LeanAttention reaches a speedup of more than 2x even at smaller contexts. This is because parallelizing only over the batch and heads (total heads $= 256 \times 4 = 1024$) does not provide sufficient work for each SM (total SMs $= 8 \times 108 = 864$) as  $1024 - (9 \times 108) = 52$ SMs remain idle in the last wave. Furthermore, FD behaves identically to vanilla FlashAttention-2, opting for a split factor of 1. In contrast, LeanAttention computes attention in fully quantized waves for all problem sizes.

To observe this effect in greater detail, we evaluate across a varying number of attention heads in \autoref{fig:8xUnitResults}(b) with a context length of 256k and batch size of 4. We observe comparable speedups to FD and FI at a smaller number of heads (64, 160). This is because at these dimensions, fixed-split is able to produce enough splits to occupy most of the SMs. We can also clearly see that FD resorts to vanilla execution when we increase the number of heads from 160 to 512. LA, on the other hand scales well as we increase the number of heads, showcasing its hardware-aware scalable execution algorithm.
With 256 heads, as we increase the batch size from 1 to 32, we can see that LA starts to outperform FA2 variants as batch size increases.

\noindent{\textbf{Effect of Head Dimension.}} \autoref{fig:headdim128} shows the speedup offered by LA for models with a LLAMA-2, Mistral and Phi3-like config with a head dimension of 128. We utilize a 128-token wide LeanTile for decomposition of each problem instead of 256.  We observe a similar trend in performance, where LA delivers a speedup of 3.5x compared to FD at 128k context length. Even at smaller context lengths of 8k tokens, we observed an improved performance of 1.34x over FD.

To summarize, LA not only outperforms FD at lower batch sizes, long context lengths, but also delivers better performance at higher batch sizes, and with higher number of attention heads. This is mainly due to the lean decomposition of the problem on the hardware compute resource. For cases where there are enough parallelizable dimensions, LA automatically generalizes to FA2-like execution. We can thus treat FA2's execution algorithm as a special case of LA, which occurs depending on the optimal grid size that LeanAttention chooses depending on the hardware resources and LeanTile dimensions. 

Fixed-split partitioning results in imbalanced workloads across the SMs, leaving many of them idle during the final stages of computation. This inefficiency makes fixed-split attention mechanisms energy-inefficient. As shown in \autoref{fig:energy}, the disparity in energy consumption between FlashDecoding, FlashInfer, and LeanAttention increases as context lengths grow over 128k. LeanAttention, with its well-balanced load partitioning strategy, ensures more consistent utilization of SMs, making it significantly more energy-efficient.

\begin{figure}[t]
\begin{center}
\centerline{\includegraphics[width=\columnwidth]{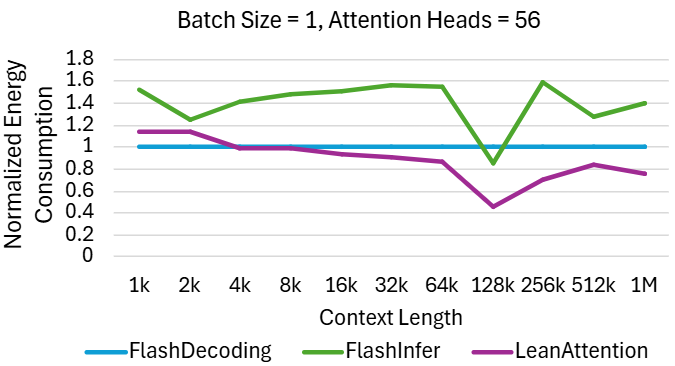}}
\vspace{-0.1in}
\caption{Ratio of Energy consumed by attention kernel to energy consumed by FlashDecoding kernel of different attention mechanisms for batch size = 1, number of heads = 56, head dimension = 64 on a single Nvidia-A100-80GB GPU as measured using NVML APIs.}
\label{fig:energy}
\end{center}
\vspace{-0.2in}

\end{figure}

\subsection{End-to-End Inference Performance}
\label{subsec:inferperf}
We measure the end-to-end inference runtime using Phi-3  Medium model (with 40 attention heads) as shown in \autoref{fig:EndToEndSpeedup} at different prompt sizes (total context = prompt tokens + tokens generated so far) with a prompt to output token ratio of 8:1. This includes the prefill-stage latency as well as the total runtime of decode-phase. LeanAttention offers a 1.12x speedup with Phi-3 Medium as compared to FlashDecoding for first 1k output tokens. However, the LA offers a higher speedup as the output tokens increase beyond 16k delivering an average of 1.73x speedup compared to FA2. As we note, the inference-level runtime improvement delivered by LA will change heavily on the number of heads, total context length, batch size, etc.

\section{Conclusion}
The attention mechanism in transformer-based language models is a slow and memory hungry process. State-of-the-art optimization mechanisms, such as FlashAttention-2, FlashDecoding, and FlashInfer, have cleverly addressed this challenge; however, they fail to adapt to the computationally distinct phases of inference. We observe that these techniques fail at parallelization along the context length dimension during the decode phase of inference, resulting in low occupancy of the underlying hardware and slower inference. As state-of-the-art models continue to push the limits on supporting increasingly long context lengths and heterogeneous context-length batching~\cite{sarathi}, the importance of optimization techniques that effectively parallelize across this dimension is becoming increasingly critical.

To address this challenge, we propose \textit{LeanAttention}, a scalable and generalized exact attention execution mechanism that ensures lower runtimes and almost 100\% hardware occupancy during attention computation irrespective of problem size. LeanAttention leverages the associative property of softmax re-scaling, treating it as a reductive operator in a "stream-K"-style partitioning of the attention mechanism. This enables an efficient parallelized execution of the attention mechanism, particularly during the decode phase of inference, which traditionally suffers from longer runtimes and critically low hardware utilization.

Our measurements indicate that LeanAttention delivers an average speedup of 1.73x over FlashDecoding, with up to 2.18x speedup for a 256k context size. Notably, in a multi-GPU execution scenario with numerous attention heads, the speedup realized by LeanAttention continues to increase as context length grows. LeanAttention intelligently utilizes the underlying hardware, enabling efficient scaling of next-generation LLMs that leverage large context lengths.

%%%%%%% -- PAPER CONTENT ENDS -- %%%%%%%%
\clearpage

%%%%%%%%% -- BIB STYLE AND FILE -- %%%%%%%%
\bibliographystyle{IEEEtranS}
\bibliography{refs}
%%%%%%%%%%%%%%%%%%%%%%%%%%%%%%%%%%%%

\end{document}